\documentclass[prd,onecolumn,
tightenlines,superscriptaddress,
amsfonts,amsmath,amssymb]{revtex4-2}
\usepackage[colorlinks,linkcolor=blue,anchorcolor=violet,citecolor=red]{hyperref}
\usepackage[pdftex]{graphicx}
\usepackage{float,color}
\usepackage{xcolor}
\usepackage{subfig}

\usepackage{algorithmic}
\usepackage{algorithm}

\usepackage{caption}
\usepackage{url}

\begin{document}
\title{Deep Learning Metric Detectors in General Relativity}

\author{Ryota Katsube}
\email{ryota.katsube.p8@dc.tohoku.ac.jp}
\affiliation{Department of Physics, Graduate School of Science, Tohoku University, Sendai 980-8578, Japan}
\author{Wai-Hong Tam}
\email{wai.tam@helsinki.fi}
\affiliation{Department of Physics, P.O. Box 64, FI-00014 University of Helsinki, Finland}
\affiliation{Department of Physics, Graduate School of Science, Nagoya
University, Chikusa, Nagoya 464-8602, Japan}
\author{Masahiro Hotta}
\email{hotta@tuhep.phys.tohoku.ac.jp}
\affiliation{Department of Physics, Graduate School of Science, Tohoku University, Sendai 980-8578, Japan}
\author{Yasusada Nambu}
\email{nambu@gravity.phys.nagoya-u.ac.jp}
\affiliation{Department of Physics, Graduate School of Science, Nagoya
University, Chikusa, Nagoya 464-8602, Japan}

\date{\today}

\begin{abstract}
We consider conceptual issues of deep learning (DL) for metric detectors using test particle geodesics in curved spacetimes.
Advantages of DL metric detectors are emphasized from a view point of general coordinate transformations.
Two given metrics (two spacetimes) are defined to be conneted by a DL isometry if their geodesic image data cannot be discriminated by any DL metric detector at any time. 
The fundamental question of when the DL isometry appears is extensively explored. If the two spacetimes connected by the DL isometry are in superposition of quantum gravity theory, the post-measurement state may be still in the same superposition even after DL metric detectors observe the superposed state. 
 We also demonstrate metric-detection DL's in 2+1 dimensional anti-de Sitter (AdS) spacetimes to estimate the cosmological constants and Brown-Henneaux charges. In the AdS/CFT correspondence dictionary, it may be expected that such metric detectors in the AdS bulk region correspond to quantum measurement devices in the CFT at the AdS boundary.
\end{abstract}
\maketitle

\section{Introduction}

How do intelligent agents perceive space and time? The agents can be human
beings, or possibly artificial intelligences (AI), which consist of
semiconductors working by some deep learning (DL) protocols. This
 question involves a lot of interdisciplinary features linking
various fields of science and philosophy. In fundamental physics, this is a quite
significant issue too. The space and time merge into a relativistic spacetime. The spacetimes are described by the general
relativity (GR) theory, and the agents can be self-reliant AI detectors based
on DL for future gravitational wave (GW) observatories. In 2015 two GW
observatories, LIGO Scientific Collaboration and Virgo Collaboration, first
observed GW spacetime deviation using large interferometers \cite{LIGO}. For
such data analyses of GW detection, several DL methods have been
already developed for extracting signals from raw data contaminated by noise
\cite{DLG1,DLG2,DLG3}. But the DL reconstruction of metric tensors
\thinspace$g_{\alpha\beta}$ for detected spacetimes has not yet been
extensively explored as a conceptual issue in GR. 
This problem is also crucial for quantum cosmology 
because the concept of observers for quantum spacetimes still remains elusive. 
Even in the AdS/CFT correspondence \cite{ADSCFT}, such a concept of observers in the bulk region should be seriously studied. The AdS/CFT correspondence implies a duality between a quantum GR theory with an anti-de Sitter (AdS) background and a conformal field theory (CFT), the dimensions of which are one dimension lower than the AdS spacetime dimensions in a holographic way. The AdS/CFT theory allows us to redefine the quantum gravity itself by using the well-defined CFT. 
 It is known in AdS/CFT that, via the famous Ryu-Takayanagi formula \cite{RT}, metric forms $g_{\alpha\beta}$ of emergent
asymptotic anti-de Sitter spacetimes can be determined from information of entanglement entropy at the AdS boundary \cite{me1,me2}. Recently, the bulk metric forms are computed using DL from certain data of conformal field theories \cite{H,H2}. But  in these previous works, explicit measurement schemes of $g_{\alpha\beta}$ are not discussed. Another interesting direction to fix the metric tensors is the correlation method of quantum field theory in curved spacetimes \cite{A1,A2, E1,P1}. By using two-point correlation functions of quantum fields in general spacetimes, the metrics are reproduced by taking derivatives of the functions with respect to the spacetime coordinates. However, in the scheme, single-shot measurements of the metrics cannot be achieved, and a lot of the same experiments are required to obtain the single metric form. In this
paper, we concentrate on the determination of the DL metric based on single-shot measurements of time-like geodesics in the spacetimes.
\begin{figure}[t]
\centering
\includegraphics[width=9cm]{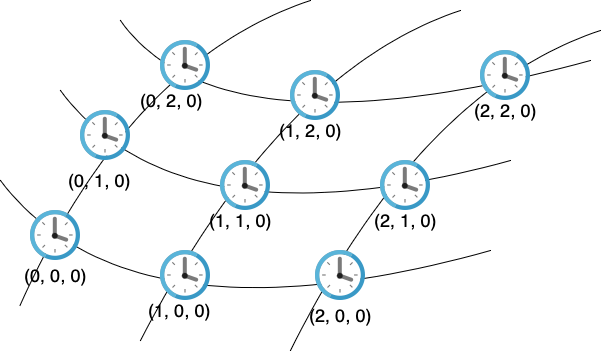}
\captionsetup{justification=raggedright}
	\caption{The clocks are distributed in the $D-1$ dimensional space, where $D=4$, and each clock is
labeled by $D-1$ variables.}
	\label{clock}
\end{figure}
\begin{figure}[t]
\captionsetup{justification=raggedright}
\centering
\includegraphics[width=9cm]{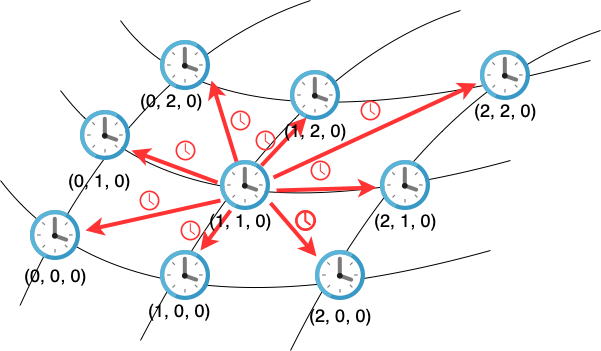}
\captionsetup{justification=raggedright}
	\caption{The clock $C_A$ is in the center sending small red clocks $C_D$, which are captured respectively by clock $C_B$ nearby.}
	\label{small_clock}
\end{figure}

In GR,  the discrimination of spacetimes is partially achieved from the mathematical
view point of Riemannian manifolds in differential geometry. Each manifold is
specified essentially by its causal structure, topological structure, and curvature data of
covariant scalars like $R_{\alpha\beta\mu\nu}R^{\alpha\beta\mu\nu}$ respectively. Here
$R_{\alpha\beta\mu\nu}$ is the Riemann curvature tensor of the spacetime.
However, from the view point of physics, such manifold properties are not
sufficient to capture the nature of curved spacetimes since all the values of
covariant scalars at a point are invariant under any general coordinate
transformation. Actually, such covariant quantities fail to describe the values
of some physical observables. For example, the energy $E$ and momentum $\vec{P}$
of black holes in asymptotically flat spacetimes change their values under the Lorentz
transformation, which is included in the general coordinate transformations. Of course, $E$ and $\vec{P}$ remain unchanged under local
coordinate transformations in a curved region, but the global coordinate
transformations in the asymptotically flat region vary the values. 

The non-covariant observables are defined by using asymptotic values of metric
tensors $g_{\alpha\beta}~$and their derivative $\partial_{\alpha}%
g_{\beta\gamma}$, or Christoffel symbols $\Gamma_{\beta\gamma}^{\alpha}$ at
spatial infinity or null future infinity. In this sense, $g_{\alpha\beta}$ and
$\Gamma_{\beta\gamma}^{\alpha}$\ in GR are partially physical, and quite different from
the ordinary gauge freedom in standard gauge theories like the non-abelian
Yang-Mills theory. In fact $g_{\alpha\beta}(x)$ and $\Gamma_{\beta\gamma
}^{\alpha}(x)$ at a point $x$ have their physical realization by constructing
a physical coordinate system in the spacetime.

For instance, the coordinate
system can be constructed by a huge number of clocks distributed in the $D-1$
dimensional space \cite{MTW}. Each clock has its own engines to accelerate and decelerate. The clock trajectories should be controlled carefully in order to avoid braiding and intersecting with each other. The clocks are
labeled by $D-1$ variables  $(x^{1},\cdots,x^{D-1})$ in order to display
their spatial positions as shown in Fig.~\ref{clock}. The displayed time $x^{0}$ on each clock panel fixes
the time coordinate at the point where the clock is located. The time $x^{0}$
does not need to be the proper time $\tau$ of the clock. Let us consider
the reconstruction of $g_{\alpha\beta}~$at a point $A$, the coordinate values of which
are denoted by $x=(x^{0},x^{1},\cdots,x^{D-1})$. \ Since $g_{\alpha\beta}$ is a
symmetric tensor in $D$ dimensions, the number of independent components of
$g_{\alpha\beta}$\ is given by $N_{g}=D(D+1)/2$ at each point. Suppose that a clock
$C_{A}$ is located at $A$. The clock $C_{A}$ emits $N_{g}$ small clocks
$C_{D}(n)$, where $n=1,2,\cdots,N_{g}$, like in Fig.~\ref{small_clock}. We suppose that the emitted clocks $C_{D}(n)$ are
propagating along geodesic lines. \ The clock $C_{D}(n)$ is captured by a
clock $C_{B}(n) $ located at $\left(x^{0}+\Delta {x}^0_{(n)},x^{1}+\Delta x_{(n)}%
^{1},\cdots,x^{D-1}+\Delta x_{(n)}^{D-1}\right)$ in a very short proper time duration 
$\Delta\tau_{(n)}$ of $C_{D}(n)$. Here we assume that the masses of all the
clocks are small enough to be ignored. The geodesics data of $C_{D}%
(n)\,$\ which are$\ $denoted by $\left(\Delta x_{(n)}^{\mu
},\Delta\tau_{(n)}\right)  $ satisfy the following equation:%
\begin{equation}
g_{\mu\nu}(x)\Delta x_{(n)}^{\mu}\Delta x
_{(n)}^{\nu}=(\Delta\tau_{(n)})^{2},
\label{e1}%
\end{equation}
where $\mu,\nu=0,1,\cdots,D-1$, and  we adopt the Einstein rule for the summation of spacetime indices, i.e., if the same index appears twice in an equation, the sum is taken. 
Introducing two $N_{g}$ dimensional column vectors defined by
\begin{align*}
&\vec g=\begin{bmatrix}
g_{00}, g_{01}, \cdots, g_{0,D-1}, g_{11}, \cdots, g_{1,D-1}, \cdots, g_{D-1,D-1}
\end{bmatrix}^T\equiv
\begin{bmatrix}
g_1,g_2,\cdots,g_{N_g}
\end{bmatrix}^T, \\
& \vec d=\begin{bmatrix}
 (\Delta\tau_{(1)})^2,(\Delta\tau_{(2)})^2,\cdots, (\Delta\tau_{(N_g)})^2
 \end{bmatrix}^T
\end{align*}
and an $N_{g}\times N_{g}$ matrix given by
$$
L=\begin{bmatrix}
\left(  \Delta x_{(1)}^0\right)\left(  \Delta x_{(1)}^0\right) &
 \left(  \Delta x_{(1)}^0\right)\left(  \Delta x_{(1)}^1\right) &
 \cdots &
\left(  \Delta x_{(1)}^{D-1}\right)\left(  \Delta x_{(1)}^{D-1}\right)\\
 \left(  \Delta x_{(2)}^0\right)\left(  \Delta x_{(2)}^0\right) &
\left(  \Delta x_{(2)}^0\right)\left(  \Delta x_{(2)}^1\right)  &
\cdots &
\left(  \Delta x_{(2)}^{D-1}\right)\left(  \Delta x_{(2)}^{D-1}\right) \\
\vdots & \vdots & \ddots & \vdots\\
\left(  \Delta x_{(N_g)}^0\right)\left(  \Delta x_{(N_g)}^0\right) &
\left(  \Delta x_{(N_g)}^0\right)\left(  \Delta x_{(N_g)}^1\right) &
\cdots & \left(  \Delta x_{(N_g)}^{D-1}\right) \left(  \Delta x_{(N_g)}^{D-1}\right)%
\end{bmatrix}  ,
$$
then Eq.\eqref{e1} is rewritten in a simple form as $L\vec{g}=\vec{d}$. By solving
this linear equation, the metric $g_{\alpha\beta}$ at $A$ is completely determined. From
this point of view, a coordinate system and its corresponding metric tensor
are physical objects which are measured and controlled in experiments. 
Beyond this old method, it is possible to consider DL methods to determine spacetime metrics $g_{\alpha\beta}$.
In this paper, we focus on fundamental issues related with the question of how the DL\ metric
detectors see the spacetime and estimate $g_{\alpha\beta}$.

\begin{figure}[b]
    \begin{tabular}{cc}
    \subfloat[Handwriting of ``2'']{
    \includegraphics[width=0.3 \linewidth]{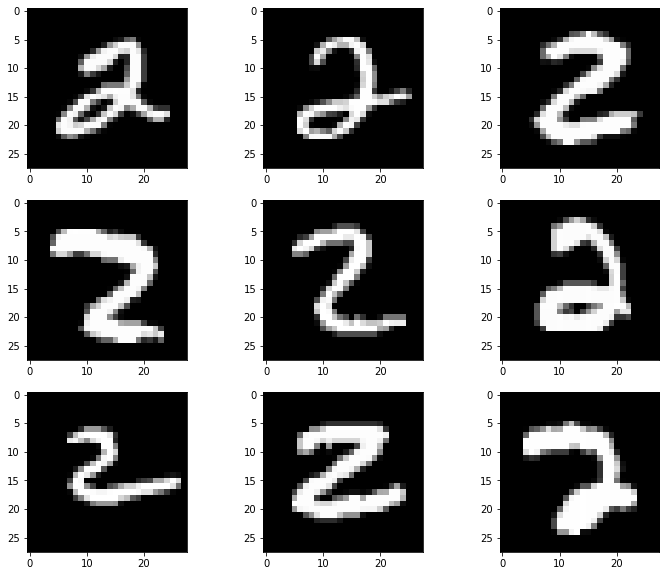}} &
    \hspace{0.5cm}
    \subfloat[Handwriting of ``7'']{
    \includegraphics[width=0.3 \linewidth]{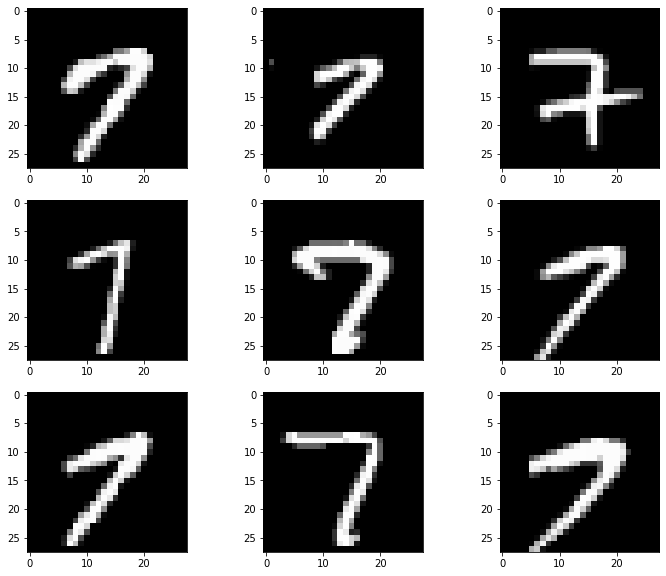}}
    \end{tabular}
	\caption{Some images in the MNIST dataset \cite{mnist}.}
	\label{fig:MNIST}
\end{figure}
\begin{figure}[tbh]
    \centering
    \includegraphics[width = 0.5\linewidth]{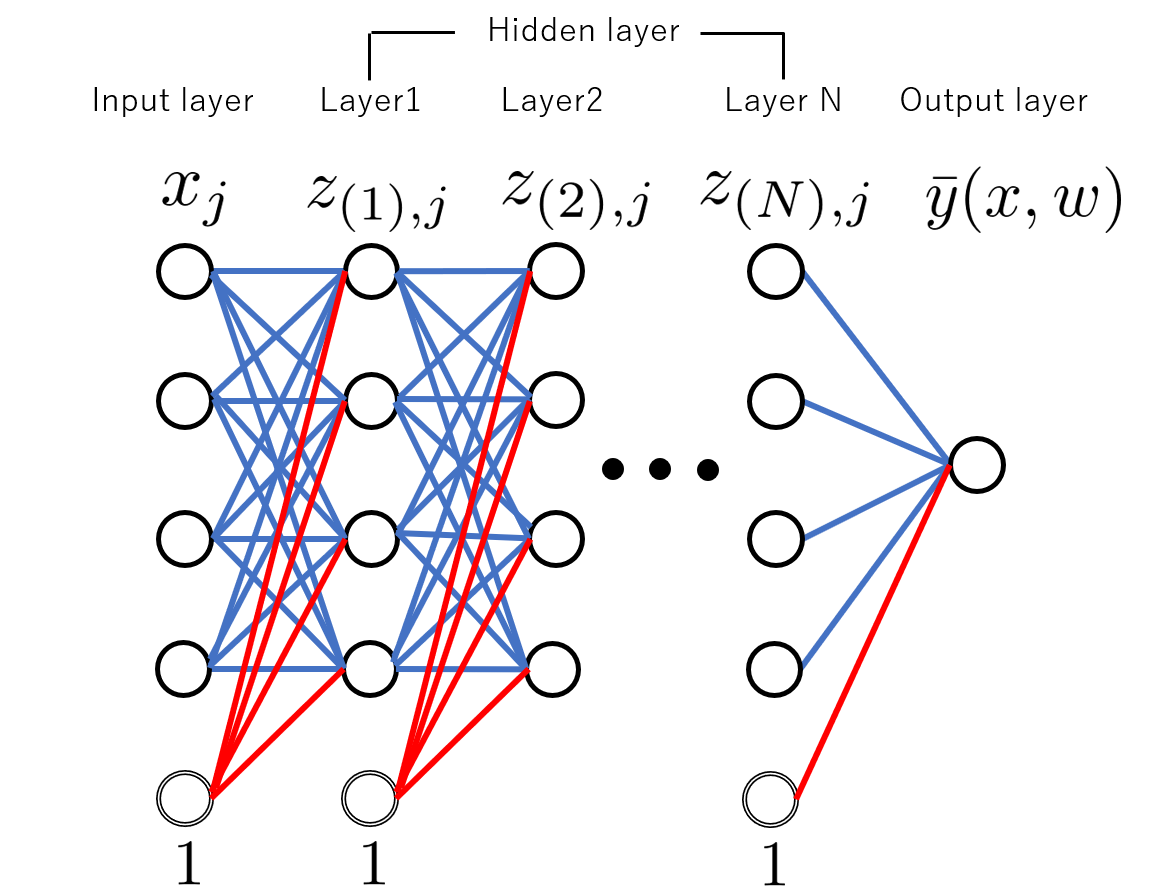}
    \captionsetup{justification=raggedright}
    \caption{The structure of a NN is depicted. The components of $x$, $z_{(l)}$ and $\bar{y}(x^{(i)},w)$ correspond to nodes in the graph, and $w_{(l),j,k}$ and $w_{(N+1),j}$  correspond to the edges which link two nodes representing variables with weight parameters associated to them. $b_{(l),j}$ corresponds to the edge which links the node representing $1$ of the $(l-1)$-th layer and the node representing $z_{(l),j}$ or $\bar{y}(x,w)$.}
    \label{fig:overviw_nn}
\end{figure}

  It can be simply said that DL is a numerical method to solve discrimination problems with high probability using large data sets \cite{Bishop} \cite{Goodfellow}. For example, suppose a handwritten digit classification task. The famous handwritten digit images data set is provided in the MNIST dataset \cite{mnist}. Some images in MNIST are presented in Fig. \ref{fig:MNIST}. The left panel presents image data indicating ``2", and the right panel presents image data indicating ``7". Of course it is easy for a human being to classify the handwritten digits. But it was difficult for AI to perform the same task in high precision before the DL method was developed. At present, some DL's succeed in clearly  discriminating ``2" and ``7" for a new input image data  after the teaching by use of the image data in Fig. \ref{fig:MNIST}. In this task, the DL's approximately reveal a relation between the new input datum and its output answer. The input datum and its corresponding answer are called  feature  vector  and  target variable, respectively. Let us denote a feature vector and a target variable of the $i$ -th datum in a data set by $x^{(i)}$ and $y^{(i)}$, respectively.  Then the task is described as finding a map $f$ that satisfies $y^{(i)} = f(x^{(i)})$ in a good approximation for any $i$. The DL methods solve this task as an optimization problem. Note that neural networks (NN)  are the most basic building block in the DL \cite{Bishop} \cite{Goodfellow}.  The NN structure is the following: First, we input a feature vector $x$ to the NN. Then, we treat an intermediate vector $z_{(1)}$, where  the subscript ``(1)" means that it is the "first" intermediate vector.  Let $x_j$ and $z_{(1),j}$ be the $j$-th component of $x$ and $z_{(1)}$, respectively. $z_{(1),j}$ is calculated by the following equation:
\begin{equation}
z_{(1),j} = \sigma\left(\sum_{k} w_{(1),j,k} x_k + b_{(1),j}\right), \notag    
\end{equation}
where  $w_{(1),j,k}$ is called weight and connects $z_{(1),j} $ to $x_k$, and the constant term $b_{(1),j}$ associated with $z_{(1),j}$ is called bias. The function $\sigma(x)$ is a non-linear function of $x$, which is called an activation function. Note that weights and biases take the same value regardless of the number $i$ of the input datum.  Similarly, we compute the $(l+1)$-th intermediate vector $z_{(l+1)}$ from the $l$-th intermediate vector $z_{(l)}$ by
\begin{equation}
z_{(l+1),j} = \sigma \left( \sum_k w_{(l+1),j,k} z_{(l),k} + b_{(l+1),j}\right), \notag    
\end{equation}
where $z_{(l+1),j}$ is the $j$-th component of the $l$-th intermediate vector, $w_{(l+1),j,k}$ is the weight which connects $z_{(l+1),j}$ to $z_{(l),k}$, and $b_{(l+1),j}$ is a constant term associated with $z_{(l+1),j}$.  This operation is repeated $N$ times. The predicted values  $\bar{y}(x,w)$ by the NN for the feature vector $x$ and given parameters $w = \{ \{w_{(l),j,k}\}_{l \ne N+1,j,k},\{w_{(N+1),j}\}_j,\{b_{(l),j}\}_{l,j}\}$ are computed as follows:
    \begin{equation}
     \bar{y}(x,w) = \sigma \left(\sum_j w_{(N+1),j} z_{(N),j} + b_{(N+1),j} \right),  \notag
    \end{equation}
where $w_{(N+1),j}$ is a weight which associates $z_{(N),j}$ with $\bar{y}(x,w)$, and $b_{(N+1),j}$ is a bias.
The model structure of NN is depicted in Fig. \ref{fig:overviw_nn}. In Fig. \ref{fig:overviw_nn}, the components of $x$, $z_{(l)}$ and $\bar{y}(x^{(i)},w)$ correspond to nodes in the graph, $w_{(l),j,k}$ and $w_{(N+1),j}$  correspond to the edges which link two nodes representing variables with weight parameters associated to them. $b_{(l),j}$ corresponds to the edge which links the node representing $1$ of the $(l-1)$-th layer and the node representing $z_{(l),j}$ or $\bar{y}(x,w)$.

The aim of the NN is to find the weight and bias parameters with which $\bar{y}(x^{(i)},w) = y^{(i)}$ approximately holds in a good precision for any $i$. To search optimized parameters, we define a loss function, which is a function that quantifies how close the NN's prediction $\bar{y}(x^{(i)},w)$ and  the actual answers $y^{(i)}$ are, by $E(w)$.  For ordinary regression problems, the mean squared errors (MSE) are often adopted as the loss functions, which are given by 
    \begin{equation}
E(w) = \frac{1}{n} \sum_{i=1}^n \left( \bar{y}(x^{(i)},w) - y^{(i)}\right)^2, \notag        
    \end{equation}
where $n$ is the total number of data. Note that if we want to solve a classification problem, in which $y^{(i)}$ is discretized,  we modify the network structure of the NN given in Fig. \ref{fig:overviw_nn}. In Fig. \ref{fig:overviw_nn}, the number of nodes of the  output layer is one. But we change that number to the number of classes we discriminate. For example, when we make a DL recognize handwritten numbers from 0 to 9, we set the number of nodes in the output layer to 10. The $c$ th  node in the output layer calculates the probability $P_c(x^{(i)},w)$ where the DL model predicts that  the $i$-th data 's target label is $c$ for given parameters $w$. The final prediction of DL,  $\bar{y}(x^{(i)},w)$ will be the class that has the highest probability $\bar{y}(x^{(i)},w) = {\rm argmax}_{c} P_c(x^{(i)},w)$.
For classification problems, another standard example of loss functions is the cross entropy given by
\begin{equation}
    E(w)=-\frac{1}{n}\sum_{i=1}^n \sum^M_{c=1} y_c(x^{(i)},w) \log P_c(x^{(i)},w),
\end{equation}
where $M$ is the total number of classes, $y_c(x^{(i)},w)$ is the target one-hot vector, which takes 1 for a target class and 0 for the rest. 
In the gradient descend method, which is the basic optimization technique, we update $w$ by
    \begin{equation}
w^{\prime} = w - \eta \frac{\partial E(w)}{\partial w} \notag,      
    \end{equation}
where $w^{\prime}$ are new parameters, and $\eta$ is a positive parameter controlling the $w$ change. Note that $\eta$ is called the learning rate and its value is given by hand.
The optimized parameters are parameters which minimize $E(w)$. By combining NN and other tools like convolution, high effcient DL's are achieved. The details are explained in Appendix A. In this paper, we study metric detectors equipped with the DL.

In the old method without DL, the hardest task to determine the spacetime metrics $g_{\mu \nu}$ is the following: not only at the single point $A$ but also at all the other
points in the spacetime, we should gather measurement data to resolve
Eq.~\eqref{e1}. This is not realistic in the experiments. Owing to this fact, a model metric tensor is often considered, which
is parametrized by $N$ real numbers 
 $\lambda=(\lambda_{1},\cdots,\lambda_{N})$: 
\begin{equation}
ds^{2}=g_{\mu\nu}(x,\lambda)dx^{\mu}dx^{\nu}.\label{e2}%
\end{equation}
The number $N$ can be set large or small depending on our interest, and a priori knowledge about the observed spacetime. The task in this case is to look for the optimized values of $\lambda$ which reproduce the actual metric tensors as precisely as possible.
As a first attempt, it may be assumed that the reference coordinate system $x$ is precisely implemented for $\lambda=(0,\cdots,0)$ such that it reproduces the planned values of $g_{\mu\nu}$ without error. But in reality or actuality, we are not capable of controlling all the coordinate clock trajectories in high precision since we often do not know the detail of the spacetime in advance. The deviation of
the trajectories can be treated as a local coordinate transformation to change
the form of $g_{\mu\nu}$. We are able to take account of such deviation
effects by considering that some of the $\lambda_{n}$'s represent the freedom of 
local coordinate transformations. Other $\lambda_{n}$'s describe the \ degrees
of freedom for both covariant quantities of the spacetime manifolds and
noncovariant physical observables like energy and momentum. In this paper, we
demonstrate a DL scheme to estimate the value of $\lambda$ based on image data
of geodesics of test particles propagating in the spacetime given by
Eq.~\eqref{e2}. It should be stressed that the test particles need not be clocks displaying their time $x^0$. In a realistic situation of astrophysics, the
particles can be dust or small stars randomly distributed in the
space with $D-1=3$. Unlike the moving clocks defining the reference coordinate system for the spacetime, we do not need to avoid tangled trajectories of the particles and their intersections. This scheme makes the $g_{\mu\nu}$
determination quite simple, admitting a model dependence of the metric form. Of course, the model should be based on the analysis of some convincing theory. For example, current gravitational wave observations adopt numerical simulation of two colliding objects like black holes and neutron stars in the GR theory with $D=4$, and make such a metric model in Eq.~\eqref{e2} as a template of emitted gravitational waves \cite{LIGO}. It may be interesting to imagine that future AI's will formulate the adopted theories for the models by themselves using unsupervised machine learning protocols based on their own experiences in the spacetime. If we are able to read out the model metric $g_{\mu\nu}(x,\lambda)$ which spontaneously emerges in the AI brains, it will provide a huge impact for various fields of science and philosophy about the space-time perception.  In this paper, using Eq.~\eqref{e2} and the geodesic equation for the spacetime, we numerically
generate image data of geodesics for fixed values of $\lambda$, and train
a DL system by these data. After that, we input different image data of
geodesics to the DL system, and estimate the value of $\lambda$ for the input data.

 Note that some DL is good at noticing that all the handwritings in the left panel of Fig. \ref{fig:MNIST}
indicate the same single symbol, i.e., ``2".  Some of the handwritings are connected by general coordinate transformations in two dimensions. Owing to the similar reason, a DL metric detector is able to automatically neglect local coordinate transformations when the aim is to get the value of physical quantities in GR. As
mentioned above, the physical quantities like energy and momentum should not
change under local coordinate transformations. But the local coordinate
transformation generates different image data of particle trajectories. Thus, it is expected that some DL's give us right answers about the physical
quantities even if the implementation of the reference coordinate system locally deviates from the anticipated
coordinate system of the original design. Hence the DL method may be an excellent
strategy to extract information of the physical observables easily from
measurement data.

In this paper we also consider another aspect of DL in GR.
For which type of metric forms does any DL metric detector fail to notice the difference owing to
fundamental reasons? At first glance, one might expect that big data of many
geodesics are so huge that the detector discriminates all different metric forms since
the form depends only on a finite number of parameters $\lambda$. Even in the case of
projected image data onto lower dimensions, a DL metric detector might also respond in a
different way for each value of $\lambda$ in Eq.~\eqref{e2} and discriminate the spacetimes. But this naive expectation is not always realized. We have counter examples as will be discussed in section 2. If any DL metric detector judges that two different metric forms are the same, we define that the metric forms are connected by a DL isometry. 
In GR, a usual isometry which preserves a metric form of a spacetime results in conservation laws of physical charges in the spacetime. Each conserved charge takes the same value in time. The DL isometry results in a generation of the same image data of particle geodesics for different metric forms. If two spacetimes connected by a DL isometry are quantum mechanically superposed, the post-measurement states are still in the same superposition even after the observation of the quantum states by the DL metric detectors. Thus the DL isometry is expected to play an interesting role in the quantum measurement theory in quantum gravity.   

In section 2, we explain that the covariant constant symmetric tensors (CCST) and the homothetic vectors
provide the DL isometry. We provide a useful integrability condition for the existence of a nontrivial CCST. 
In section 3, we consider a DL for estimating 
cosmological constants for anti-deSitter (AdS) spacetimes. We
provide a simple proof that a nontrivial DL isometry does not exist in this case. 
It is also demonstrated that a DL can identify modified image data connected by local
coordinate transformations and estimate the same value of the cosmological
constant. In section 4, we demonstrate a DL for the estimation of  Brown-Henneaux charges of AdS metrics. The results are crucial for metric detection of the bulk part in the AdS/CFT correspondence. Based on our argument, it is natural to consider that the metric detectors correspond to quantum measurement devices for conformal fields at the AdS boundary. This will enable us to profoundly understand what a measurement in the AdS/CFT correspondence is. In section 5, we summarize our results.

\section{Deep Learning Isometry}

In this section, we define and explore the DL isometry in the presence of
which any DL metric detector fails to distinguish different curved spacetimes. Let us consider
a $D$ dimensional general metric form:%
$$
ds^{2}=g_{\mu\nu}(x)dx^{\mu}dx^{\nu},
$$
where $x=(x^{0},x^{1},\cdots,x^{D-1})$. The Christoffel symbols $\Gamma
_{\nu\nu^{\prime}}^{\mu}(x)$ are defined as%
\begin{equation}
\Gamma_{\nu\nu^{\prime}}^{\mu}(x)=\frac{1}{2}g^{\mu\mu^{\prime}}(x)\left(
\frac{\partial g_{\mu^{\prime}\nu^{\prime}}}{\partial x^{\nu}}(x)+\frac
{\partial g_{\nu\mu^{\prime}}}{\partial x^{\nu^{\prime}}}(x)-\frac{\partial
g_{\nu\nu^{\prime}}}{\partial x^{\mu^{\prime}}}(x)\right)  .\label{e7}%
\end{equation}
By using $\Gamma_{\nu\nu^{\prime}}^{\mu}(x)$, the geodesic equation reads%
\begin{equation}
\frac{d^{2}x^{\mu}}{d\tau^{2}}+\Gamma_{\nu\nu^{\prime}}^{\mu}\left(
x(\tau)\right)  \frac{dx^{\nu}}{d\tau}\frac{dx^{\nu^{\prime}}}{d\tau
}=0,\label{e8}%
\end{equation}
for a particle trajectory $x(\tau)=(x^{0}(\tau),\cdots,x^{D-1}(\tau))$ and its
proper time $\tau$. The time-like geodesics also satisfy%
$$
g_{\nu\nu^{\prime}}(x\left(  \tau\right)  )\frac{dx^{\nu}}{d\tau}\frac
{dx^{\nu^{\prime}}}{d\tau}=-1.
$$
The above equation fixes the unit of proper time.

Let us define geodesic image data $\mathcal{I}$ for DL as follows. First, 
$\mathcal{I}$ is a subset of $\mathcal{G}$, and $\mathcal{G}$\ is defined as
geodesic image data of the spacetime such that%
$$
\mathcal{G}=\left\{ \text{trajectory of}~ \left(  x^{0}\left(  \tau\right)  ,\cdots,x^{D-1}\left(
\tau\right)  \right)  | \tau_i \leq \tau \leq \tau_f ,\left(  x^{\mu}(0),\frac{dx^{\mu}}{d\tau}(0)\right)
\in\mathcal{D}\right\},
$$
where $\tau_i$ ($\tau_f$) is the start point time (the end point time) of the trajectories satisfying $0 < \tau_i < \tau_f $, the initial conditions $\left(  x^{\mu}(0),\frac{dx^{\mu}}{d\tau
}(0)\right)  $ of geodesics $x^{\mu}(\tau)$\ belong to a domain $\mathcal{D}
$ that we fix, depending on our interest. Usually, $\mathcal{D}$ is given by the
entire region spanned by the coordinate system. Here, the explicit values of $\tau_i$ and $\tau_f$ are fixed such that no geodesic trajectory collides with  singularities of the spacetime. We do not measure the proper time $\tau$ of each particle in this scheme. Thus, in the image data $\mathcal{G}$ and $\mathcal{I}$, the information about the parameter $\tau$ for each trajectory is lost. Only the information about forms of non-parametrized trajectories in the spacetime is recorded. One example of $\mathcal{I}$ is a data set of a finite positive number $M$ of geodesics:%
$$
\mathcal{I}_{M}=\left\{ \text{trajectory of}~   \left(  x_{(j)}^{0}\left(  \tau\right)  ,\cdots
,x_{(j)}^{D-1}\left(  \tau\right)  \right)  |\tau_i \leq \tau \leq \tau_f, j=1,\cdots,M\right\} \subset{\mathcal{G}} .
$$
The projected image data of $\mathcal{I}_M$  onto the $(x^1,x^2)$ plane is another example:
$$
 \mathcal{I}_{M}^{(1,2)}=\left\{ \text{trajectory of}~   \left(  x_{(j)}^{1}\left(  \tau\right), x_{(j)}^{2}\left(  \tau\right)  \right)  |\tau_i \leq \tau \leq \tau_f, j=1,\cdots,M\right\}. 
$$

Let us consider two different metrics $g_{\mu\nu}$ and $\bar{g}_{\mu\nu}$. If the
geodesic image data generated by the metrics coincide, i.e., $\mathcal{I=\bar
{I}}$, any DL using $\mathcal{I}$ always fails to discriminate $g_{\mu\nu}$
and $\bar{g}_{\mu\nu}$. When $\mathcal{I=\bar{I}}$ holds, we define that
$g_{\mu\nu}$ and $\bar{g}_{\mu\nu}$ are connected by a DL isometry for
$\mathcal{I}$. It is also defined that the two spacetimes are connected by the DL isometry. Note that $\mathcal{G=\bar{G}}$ implies $\mathcal{I=\bar{I}}$
for any $\mathcal{I}$. If $\mathcal{G=\bar{G}}$\ holds, we define that the DL
isometry is maximal for $g_{\mu\nu}$ and $\bar{g}_{\mu\nu}$.
Let us suppose that $\bar{g}_{\mu\nu}$ is given in a coodinate system $\bar
{x}^{\mu}$ as%
$$
ds^{2}=\bar{g}_{\mu\nu}(\bar{x})d\bar{x}^{\mu}d\bar{x}^{\nu}.
$$
In this coordinate system, the geodesics equation is given by%
\begin{equation}
\frac{d^{2}\bar{x}^{\mu}}{d\tau^{2}}+\bar{\Gamma}_{\nu\nu^{\prime}}^{\mu
}\left(  \bar{x}(\tau)\right)  \frac{d\bar{x}^{\nu}}{d\tau}\frac{d\bar{x}%
^{\nu^{\prime}}}{d\tau}=0,\label{e10}%
\end{equation}
where%
\begin{equation}
\bar{\Gamma}_{\nu\nu^{\prime}}^{\mu}(\bar{x})=\frac{1}{2}\bar{g}^{\mu
\mu^{\prime}}(\bar{x})\left(  \partial_{\nu}\bar{g}_{\mu^{\prime}\nu^{\prime}%
}(\bar{x})+\partial_{\nu^{\prime}}\bar{g}_{\nu\mu^{\prime}}(\bar{x}%
)-\partial_{\mu^{\prime}}\bar{g}_{\nu\nu^{\prime}}(\bar{x})\right)
.\label{e9}%
\end{equation}
Then $\mathcal{G=\bar{G}}$ implies the function form of $\bar{\Gamma}_{\nu
\nu^{\prime}}^{\mu}(x)$ in Eq. \eqref{e9} is equal to the function form of
$\Gamma_{\nu\nu^{\prime}}^{\mu}\left(  x\right)  $ in Eq.\eqref{e7}. Thus the
nessesary condition of the maximal DL isometry is given by
\begin{equation}
\Gamma_{\nu\nu^{\prime}}^{\mu}\left(  x\right)  =\bar{\Gamma}_{\nu\nu^{\prime
}}^{\mu}(x).\label{e11}%
\end{equation}

Next let us consider the metric $g_{\mu\nu}(x,\lambda)$ parametrized by
$\lambda=(\lambda_{1},\cdots,\lambda_{N})$\ in Eq.~\eqref{e2}. For an
infinitesimal deviation $\delta\lambda=(\delta\lambda_{1},\cdots,\delta
\lambda_{N})$ of $\lambda$, the metric deviation is computed as
$$
\delta g_{\mu\nu}(x,\lambda)=\sum_{n=1}^{N}\delta\lambda_{n}\frac{\partial
g_{\mu\nu}}{\partial\lambda_{n}}(x,\lambda).
$$
Then, the maximal DL isometry condition in Eq.~\eqref{e11} is reduced into
\begin{equation}
\delta\Gamma_{\beta\gamma}^{\alpha}=\frac{1}{2}g^{\alpha\mu}\left(
\nabla_{\beta}\delta g_{\mu\gamma}+\nabla_{\gamma}\delta g_{\mu\beta}%
-\nabla_{\mu}\delta g_{\beta\gamma}\right)  =0.\label{1}%
\end{equation}
Since the following relation holds
$$
\nabla_{\alpha}\delta g_{\beta\gamma}=g_{\beta\mu}\delta\Gamma_{\alpha\gamma
}^{\mu}+g_{\gamma\mu}\delta\Gamma_{\alpha\beta}^{\mu},
$$
Eq.~\eqref{1} is rewritten as
\begin{equation}
\nabla_{\alpha}\delta g_{\beta\gamma}=0,\label{2}%
\end{equation}
where the covariant derivative $\nabla_{\alpha}$ is defined by using
$g_{\alpha\beta}(x,\lambda)$. In general, a symmetric tensor $T_{\beta\gamma
}(x)$ satisfying $\nabla_{\alpha}T_{\beta\gamma}=0$ is referred to as a
covariant constant symmetric tensor (CCST) associated with $g_{\alpha\beta
}$. Thus Eq.~\eqref{2} means that $g_{\alpha\beta}(x,\lambda)$ and
$g_{\alpha\beta}(x,\lambda)+\delta g_{\alpha\beta}(x,\lambda)$ can be connected
by the maximal DL isometry if and only if $\delta g_{\alpha\beta}(x)$ is a CCST
associated with $g_{\alpha\beta}$. Since $\nabla_{\alpha}%
g_{\beta\gamma}=0$ holds for an arbitrary metric $g_{\alpha\beta}$, a trivial
solution of Eq.~\eqref{2} is given by%
\begin{equation}
\delta g_{\alpha\beta}(x,\lambda)=\delta C\left(  \lambda\right)
g_{\alpha\beta}(x,\lambda),\label{4}%
\end{equation}
where $\delta C(\lambda)$ is an infinitesimal conformal factor independent of
$x$. 

If the deviation $\delta g_{\alpha\beta}$ is generated by a Lie transport
of $g_{\alpha\beta}$ associated with a vector field $\epsilon^{\mu}~$such that
$\delta g_{\alpha\beta}=\nabla_{\alpha}\epsilon_{\beta}+\nabla_{\beta}%
\epsilon_{\alpha}$, then Eq.~\eqref{2} is recasted into
\begin{equation}
\nabla_{\alpha}\left(  \nabla_{\beta}\epsilon_{\gamma}+\nabla_{\gamma}%
\epsilon_{\beta}\right)  =0.\label{3}%
\end{equation}
If $\epsilon^{\mu}$\ obeys the following equation with a scalar function $\delta k(\lambda)$,
\begin{equation}
\nabla_{\alpha}\epsilon_{\beta}+\nabla_{\beta}\epsilon_{\alpha}=\delta
k\left(  \lambda\right)  g_{\alpha\beta},\label{6}%
\end{equation}
then $\delta k\left(  \lambda\right)  $ is computed as
\begin{equation}
\delta k\left(  \lambda\right)  =\frac{2}{D}\nabla_{\mu}\epsilon^{\mu
}\label{5}%
\end{equation}
by taking the trace in Eq.~\eqref{6}. Thus, $\epsilon^{\mu}$ is a conformal
Killing vector with Eq.~\eqref{5}. In general, a vector field $V^{\mu}$
satisfing
$$
\nabla_{\beta}V_{\gamma}+\nabla_{\gamma}V_{\beta}=k\,g_{\beta\gamma}%
$$
with a constant $k$ is called a homothetic vector. Thus, the vector field
$\epsilon^{\mu}$ satisfying Eq.~\eqref{6} is a homothetic vector, and the
infinitesimal coordinate transformation $x^{\prime\mu}=x^{\mu}+\epsilon^{\mu}$
yields a DL isometry.

Next let us provide some examples of DL isometries:
\begin{itemize}
\item[(1)] Consider a metric in the following form:
\begin{equation}
ds^2=h_{ab}(x^{0},\cdots,x^{S},\lambda)dx^{a}%
dx^{b}+H_{AB}(X^{S+1},\cdots,X^{D-1})dX^{A}dX^{B},\label{7}%
\end{equation}
where the indices $a,b$ run from $0$ to some integer $S$ smaller than $D-1$,
and the indices $A,B$ run from $S+1$ to $D-1$. Then let us consider projected image
data $\mathcal{I}^{(S+1,\cdots, D-1)}$ into the subspace spanned by $(X^{S+1},\cdots,X^{D-1})$ for the
metric in Eq.~\eqref{7} with $\lambda=(\lambda_{1},\cdots,\lambda_{N}%
)\neq(0,\cdots,0)$. Another image data $\mathcal{\bar{I}}$ is generated by the
metric in Eq.~\eqref{7} with $\bar{\lambda}=(0,\cdots,0)$. Since the function
$H_{AB}(X^{S+1},\cdots,D^{D-1})$ does not have $\lambda$ dependence,
$\mathcal{I}^{(S+1,\cdots, D-1)}=\mathcal{\bar{I}}^{(S+1,\cdots, D-1)}$ is satisfied. This provides a trivial example of DL
isometry between $g_{\mu\nu}(x,\lambda)$ and $g_{\mu\nu}(x,\bar{\lambda})$ for $\mathcal{I}^{(S+1,\cdots, D-1)}$.
\item[(2)] Consider a metric depending on $\lambda=(\lambda_{1},\lambda_{2})$ in the
following form:
\begin{equation}
ds^2 =e^{\lambda_{1}}h_{ab}(x^{0},\cdots
,x^{S})dx^{a}dx^{b}+e^{\lambda_{2}}H_{AB}(X^{S+1},\cdots,X^{D-1})dX^{A}%
dX^{B}.\label{8}%
\end{equation}
Then it is easy to check $\Gamma_{BC}^{a}=\Gamma_{bC}^{A}=\Gamma_{Bc}^{a}=\Gamma_{bc}^{A}=0 $
and
\begin{align*}
\Gamma_{bc}^{a}  & =\frac{1}{2}h^{ad}\left(  \partial_{b}h_{dc}+\partial
_{c}h_{db}-\partial_{d}h_{bc}\right)  ,\\
\Gamma_{BC}^{A}  & =\frac{1}{2}H^{AD}\left(  \partial_{B}H_{DC}+\partial
_{C}H_{DB}-\partial_{D}H_{BC}\right)  .
\end{align*}
This means that all the components of $\Gamma_{\beta\gamma}^{\alpha}$ are independent of $\lambda$. For $\lambda\neq\bar{\lambda}$,  there exists
 a DL isometry between $g_{\mu\nu}(x,\lambda)$ and $g_{\mu\nu}%
(x,\bar{\lambda})$. In this case, a CCST $\delta g_{\mu\nu}$ satisfying
$\nabla_{\alpha}\delta g_{\mu\nu}=0$ is given by the following form:
$$
\delta g_{\mu\nu}dx^{\mu}dx^{\nu}=\delta\lambda_{1} e^{\lambda_{1}}h_{ab}(x^{0},\cdots
,x^{S})dx^{a}dx^{b}+\delta\lambda_{2} e^{\lambda_{2}} H_{AB}(X^{S+1},\cdots,X^{D-1}%
)dX^{A}dX^{B}.
$$
\item[(3)] Consider a metric dependent on $\lambda$ in the following form:
\begin{eqnarray}
ds^2 =e^{k x^{0}+\lambda} \left(  -\left(
dx^{0}\right)  ^{2}+H_{AB}(X^1 \cdots,X^{D-1})dX^{A}dX^{B}\right),
\label{e200}
\end{eqnarray}
where $k$ is a constant, and the indices $A,B$ run from $1$ to $D-1$. This metric has a homothetic
vector $\epsilon^{\mu}=(\epsilon^0,\epsilon^1,\cdots,\epsilon^{D-1})^T \propto\left(  1,0,\cdots,0\right)  ^{T}$. In fact, let
us consider a coordinate transformation given by $x^{0}\rightarrow
x^{0}+\delta c$ with a constant $\delta c$. Then, the metric $g_{\mu\nu
}(x,\lambda)~$becomes $e^{k\delta c}\,g_{\mu\nu}(x,\lambda)$. \ Thus
$\delta g_{\mu\nu}=\nabla_{\mu}\epsilon_{\nu}+\nabla_{\nu}\epsilon_{\mu}$
becomes $\delta k\, g_{\mu\nu}$ with a constant $\delta k$, and satisfies $\nabla_{\alpha
}\delta g_{\beta\gamma}=0$. The change of metric can be described by $\delta \lambda =k\delta c$. In this case, it turns out that $g_{\alpha\beta}(x,\lambda)$
and $g_{\alpha\beta}(x,\lambda)+\delta g_{\alpha\beta}(x,\lambda)$ are connected by the maximal DL isometry even after taking account of the causal  structure of the particle trajectories which appears in the image data.
\end{itemize}
The above examples (1) and (2) are constructed by the reducible spacetimes
described by the following metric form in a coordinate system:
\begin{equation}
ds^{2}=h_{ab}(x^{0},\cdots,x^{S})dx^{a}dx^{b}+H_{AB}(X^{S+1},\cdots
,X^{D-1})dX^{A}dX^{B}.\label{10}%
\end{equation}
It was proven \cite{E} that any spacetime allowing the existence of CCST
should obey Eq.~\eqref{10} in some coordinate system. 
The metric form of example (3) can be also interpreted as that in Eq.~\eqref{10} by taking $S=D-1$.
If some direction of
$\delta\lambda=(\delta\lambda_{1},\cdots,\delta\lambda_{N})$ is a DL isometry
freedom, DL cannot discriminate the two metrics $g_{\mu\nu}(x,\lambda)$ and
$g_{\mu\nu}(x,\lambda)+\delta g_{\mu\nu}(x,\lambda)$ and the DL efficiency of
spacetime recognition is completely lost for the direction $\delta\lambda$.
The simplest way to avoid such a DL isometry in any coordinate system is just to make $g^{\mu\nu}\delta
g_{\mu\nu}$ depend on $x$. This is because any DL isometry requires $g^{\mu\nu}\delta g_{\mu\nu
}=const$. In fact, $\partial_{\alpha}\left(  g^{\mu\nu}\delta g_{\mu\nu
}\right)  =$ $\nabla_{\alpha}\left(  g^{\mu\nu}\delta g_{\mu\nu}\right)
=g^{\mu\nu}\nabla_{\alpha}\delta g_{\mu\nu}=0$ is satisfied for any CCST
$\delta g_{\mu\nu}$.

Next let us consider integrability condisitions of the CCST equation
$\nabla_{\alpha}\delta g_{\beta\gamma}=0$. Since we have  $D$
simultaneous equations with $\alpha=0,\cdots,D-1$ for $\delta g_{\beta\gamma}%
$, the differential equation cannot be integrated for an arbitrary initial condition, i.e., a nontrivial 
CCST does not always exist. If we have a solution of the equation,%
$$
\left[  \nabla_{\mu},\nabla_{\nu}\right]  \delta g_{\beta\gamma}=\nabla_{\mu
}\left(  \nabla_{\nu}\delta g_{\beta\gamma}\right)  -\nabla_{\nu}\left(
\nabla_{\mu}\delta g_{\beta\gamma}\right)  =0
$$
always holds since $\nabla_{\nu}\delta g_{\beta\gamma}=\nabla_{\mu}\delta
g_{\beta\gamma}=0$ holds. From the identity equation  $\left[  \nabla_{\mu}%
,\nabla_{\nu}\right]  \delta g_{\beta\gamma}=-R^{\alpha}{}_{\beta\mu\nu}\delta
g_{\alpha\gamma}-R^{\alpha}{}_{\gamma\mu\nu\delta}\, g_{\alpha\beta}$, the above
equation leads to%
\begin{equation}
R^{\alpha}{}_{\beta\mu\nu}\,\delta g_{\alpha\gamma}+R^{\alpha
}{}_{\gamma\mu\nu}\,\delta g_{\alpha\beta}=0.\label{11}%
\end{equation}
Now let us show that if Eq.~\eqref{11} holds for a symmetric tensor $\delta
g_{\beta\gamma}(x)~$in the spacetime, $\delta g_{\beta\gamma}(x)$ is computed
using the following integration formula:%
\begin{equation}
\delta g_{\beta\gamma}(x)=\delta g_{\beta\gamma}(0)+x^{\mu}\int_{0}^{1}%
\Gamma_{\mu\beta}^{\alpha}(ux)\delta g_{\alpha\gamma}(ux)du+x^{\mu}\int
_{0}^{1}\Gamma_{\mu\gamma}^{\alpha}(ux)\delta g_{\alpha\beta}(ux)du.\label{12}%
\end{equation}
When the point $\left(  x^{0},\cdots,x^{D-1}\right)  $ is very close to the
origin of the coordinate system, $\left(  0,\cdots,0\right)  $, the formula
becomes%
\begin{equation}
\delta g_{\beta\gamma}(x)\approx\delta g_{\beta\gamma}(0)+x^{\mu}\,\Gamma
_{\mu\beta}^{\alpha}(0)\delta g_{\alpha\gamma}(0)+x^{\mu}\,\Gamma_{\mu\gamma
}^{\alpha}(0)\delta g_{\alpha\beta}(0).\label{13}%
\end{equation}
Equation \eqref{12} implies that the value of $\delta g_{\beta\gamma}$ at a point $x$ can
be recursively determined by the values of $\delta g_{\beta\gamma}$ at points
$ux^{\mu}$ with $0\leq u<1$, i.e., more closer points to the origin. Hence, the
equation $\nabla_{\alpha}\delta g_{\beta\gamma}=0$ is integrable if
Eq.~\eqref{11} holds at any point.

The proof of Eq.~\eqref{12} is the following. The CCST equation is rewritten as
\begin{equation}
\partial_{\mu}\delta g_{\beta\gamma}=A_{\mu,\beta\gamma} 
,\quad
A_{\mu,\beta\gamma}=\Gamma_{\mu\beta}^{\alpha}\delta g_{\alpha\gamma}%
+\Gamma_{\mu\gamma}^{\alpha}\delta g_{\alpha\beta}.
\label{eq:A}
\end{equation}
Then the integrability condition for $A_\mu$ is given by
\begin{equation}
\partial_{\mu}A_{\nu,\beta\gamma}-\partial_{\nu}A_{\mu,\beta\gamma
}=0.\label{15}%
\end{equation}
From this equation, it is possible to show that the following circular
integral vanishes for  an arbitrary close path of integration:
$$
\oint\partial_{\mu}\delta g_{\beta\gamma}(x^{\prime})dx^{\prime\mu}=%
\oint A_{\mu,\beta\gamma}(x^{\prime})dx^{\prime\mu}=0.
$$
 This means that the metric difference
\begin{equation}
\delta g_{\beta\gamma}(x)-\delta g_{\beta\gamma}(0)=\int_{0}^{x}\partial_{\mu
}\delta g_{\beta\gamma}(x^{\prime})dx^{\prime\mu}=\int_{0}^{1}A_{\mu
,\beta\gamma}(x^{\prime}(u))\frac{dx^{\prime\mu}}{du}du\label{17}%
\end{equation}
is computed independently of its integration path $x^{\prime\mu}=x^{\prime\mu}(u)$, which
satisfies $\left(  x^{\prime0}(0),\cdots,x^{\prime D-1}(0)\right)  =\left(
0,\,\cdots,0\right)  \ $and $\left(  x^{\prime0}(1),\cdots,x^{\prime
D-1}(1)\right)  =\left(  x^{0},\,\cdots,x^{D-1}\right)  $. Then Eq.~\eqref{12}
is derived by taking $x^{\prime\mu}(u)=u\,x^{\mu}$ in Eq.~\eqref{17} and
substituting $
A_{\mu,\beta\gamma}=\Gamma_{\mu\beta}^{\alpha}\,\delta g_{\alpha\gamma}%
+\Gamma_{\mu\gamma}^{\alpha}\,\delta g_{\alpha\beta}
$. The remaining task for the proof is to show that
Eq.~\eqref{15} is equal to Eq.~\eqref{11}. Let us write  Eq.~\eqref{15} as%
\begin{align*}
& \partial_{\mu}\Gamma_{\nu\beta}^{\alpha}\,\delta g_{\alpha\gamma}%
+\partial_{\mu}\Gamma_{\nu\gamma}^{\alpha}\,\delta g_{\alpha\beta}+\Gamma
_{\nu\beta}^{\alpha}\,\partial_{\mu}\delta g_{\alpha\gamma}+\Gamma_{\nu\gamma
}^{\alpha}\,\partial_{\mu}\delta g_{\alpha\beta}\\
& -\partial_{\nu}\Gamma_{\mu\beta}^{\alpha}\,\delta g_{\alpha\gamma}%
-\partial_{\nu}\Gamma_{\mu\gamma}^{\alpha}\,\delta g_{\alpha\beta}-\Gamma
_{\mu\beta}^{\alpha}\,\partial_{\nu}\delta g_{\alpha\gamma}-\Gamma_{\mu\gamma
}^{\alpha}\,\partial_{\nu}\delta g_{\alpha\beta}\\
& =0.
\end{align*}
Using Eq.~\eqref{eq:A}, it turns out that the above equation becomes%
\begin{align*}
& \left(  \partial_{\mu}\Gamma_{\nu\beta}^{\mu^{\prime}}-\partial_{\nu}%
\Gamma_{\mu\beta}^{\mu^{\prime}}+\Gamma_{\nu\beta}^{\alpha}\Gamma_{\mu\alpha
}^{\mu^{\prime}}-\Gamma_{\mu\beta}^{\alpha}\Gamma_{\nu\alpha}^{\mu^{\prime}%
}\right)  \delta g_{\mu^{\prime}\gamma}\\
& +\left(  \partial_{\mu}\Gamma_{\nu\gamma}^{\mu^{\prime}}-\partial_{\nu
}\Gamma_{\mu\gamma}^{\mu^{\prime}}+\Gamma_{\nu\gamma}^{\alpha}\Gamma
_{\mu\alpha}^{\mu^{\prime}}-\Gamma_{\mu\gamma}^{\alpha}\Gamma_{\nu\alpha}%
^{\mu^{\prime}}\right)  \delta g_{\mu^{\prime}\beta}\\
& =0,
\end{align*}
and this is precisely equal to Eq.~\eqref{11}.

 Note that there exists a trivial solution of Eq.~\eqref{11} as $\delta
g_{\beta\gamma}(x)=Cg_{\beta\gamma}(x)$ with a constant $C$. This fact is
easily proven by using  $R_{\gamma\beta\mu\nu}=-R_{\beta\gamma\mu\nu}$. It is also possible to
consider the condition for the CCST $\delta g_{\beta\gamma}(x)$ which is not
proportional to $g_{\beta\gamma}(x)$. For such a $\delta g_{\beta\gamma}(x)$,
Eq.~\eqref{11} yields an infinite number of constraints of the initial value
$\delta g_{\beta\gamma}(0)$ at the origin. The first constraint equation is
obtained by taking $x=0$ in Eq.~\eqref{11} as
$$
R^{\alpha}{}_{\beta\mu\nu}(0)\delta g_{\alpha\gamma}(0)+R^{\alpha}{}_{\gamma\mu\nu}%
(0)\delta g_{\alpha\beta}(0)=0.
$$
Let us take a partial derivative of Eq.~\eqref{11} with respect to
$x^{\mu^{\prime}}$ and substitute $x=0$ into the obtained equation. The the
following equation holds:%
$$
\partial_{\mu^{\prime}}R^{\alpha}{}_{\beta\mu\nu}(0)\delta g_{\alpha\gamma
}(0)+\partial_{\mu^{\prime}}R^{\alpha}{}_{\gamma\mu\nu}(0)\delta g_{\alpha\beta
}(0)+R^{\alpha}{}_{\beta\mu\nu}(0)\partial_{\mu^{\prime}}\delta g_{\alpha\gamma
}(0)+R^{\alpha}{}_{\gamma\mu\nu}(0)\partial_{\mu^{\prime}}\delta g_{\alpha\beta
}(0)=0.
$$
Using Eq.~\eqref{eq:A}, this equation becomes the following
second constraint equation for $\delta g_{\beta\gamma}(0)$:%
\begin{align*}
& \partial_{\mu^{\prime}}R^{\alpha}{}_{\beta\mu\nu}(0)\delta g_{\alpha\gamma
}(0)+\partial_{\mu^{\prime}}R^{\alpha}{}_{\gamma\mu\nu}(0)\delta g_{\alpha\beta
}(0)\\
& +R^{\alpha}{}_{\beta\mu\nu}(0)\left(  \Gamma_{\mu^{\prime}\alpha}%
^{\alpha^{\prime}}(0)\delta g_{\alpha^{\prime}\gamma}(0)+\Gamma_{\mu^{\prime
}\gamma}^{\alpha^{\prime}}(0)\delta g_{\alpha^{\prime}\alpha}(0)\right) \\
& +R^{\alpha}{}_{\gamma\mu\nu}(0)\left(  \Gamma_{\mu^{\prime}\alpha}%
^{\alpha^{\prime}}(0)\delta g_{\alpha^{\prime}\beta}(0)+\Gamma_{\mu^{\prime
}\beta}^{\alpha^{\prime}}(0)\delta g_{\alpha^{\prime}\alpha}(0)\right) \\
& =0.
\end{align*}
The above equation can be recasted into $
\nabla_{\mu^{\prime}}R^{\alpha}{}_{\beta\mu\nu}(0)\delta g_{\alpha\gamma
}(0)+\nabla_{\mu^{\prime}}R^{\alpha}{}_{\gamma\mu\nu}(0)\delta g_{\alpha\beta
}(0)=0
$, 
which can be also derived simply by taking a covariant derivative of
Eq.~\eqref{11} and using $\nabla_{\alpha}\delta g_{\beta\gamma}=0$.

Taking higher derivatives of Eq.~\eqref{11} generates stringent constraint
equations for $\delta g_{\beta\gamma}(0)$. Thus nontrivial solutions of
$\delta g_{\beta\gamma}(0)$, which are not proportional to $g_{\beta\gamma
}(0)$,\ are allowed only for very specific spacetimes. Note that the
integrability condition in Eq.~\eqref{11}  is covariant under coordinate
transformations. Hence, in order to explore CCST's, we do not need to find an appropriate 
coordinate system in which the metric is given by Eq.~\eqref{10}. This provides
a new method different from the eigenvalue equation method adopted in
\cite{E}. Actually, in section 3, we use Eq.~\eqref{11} to demonstrate a simple
covariant proof that de Sitter (dS) and AdS spacetimes do not have nontrivial CCST's.

\begin{figure}[]
    \centering
    \subfloat[\centering $g_{\mu \nu}$]{{\includegraphics[width=5cm]{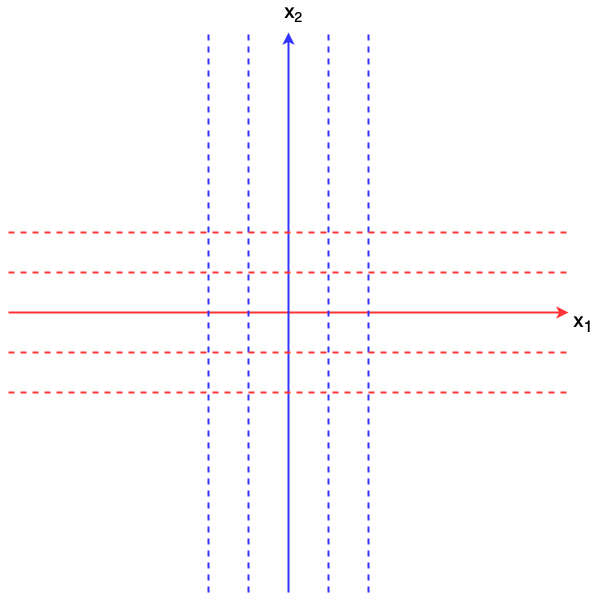}\label{metric_g} }}%
    \qquad
    \subfloat[\centering $\bar{g}_{\mu \nu}$]{{\includegraphics[width=5cm]{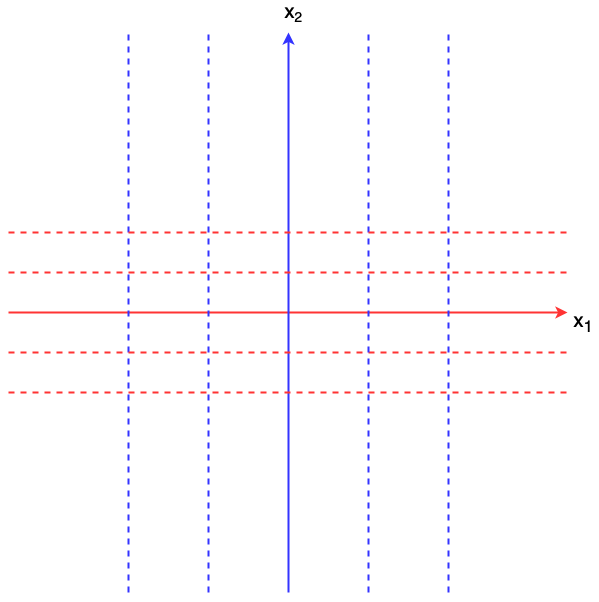}\label{metric_gbar} }} %
    \captionsetup{justification=raggedright}
    \caption{Two different coordinate systems which provide different constant metric tensors of the same flat spacetime. }%
    \label{}%
\end{figure}

\begin{figure}[]
    \centering
    \subfloat[\centering $g_{\mu \nu}$]{{\includegraphics[width=6cm]{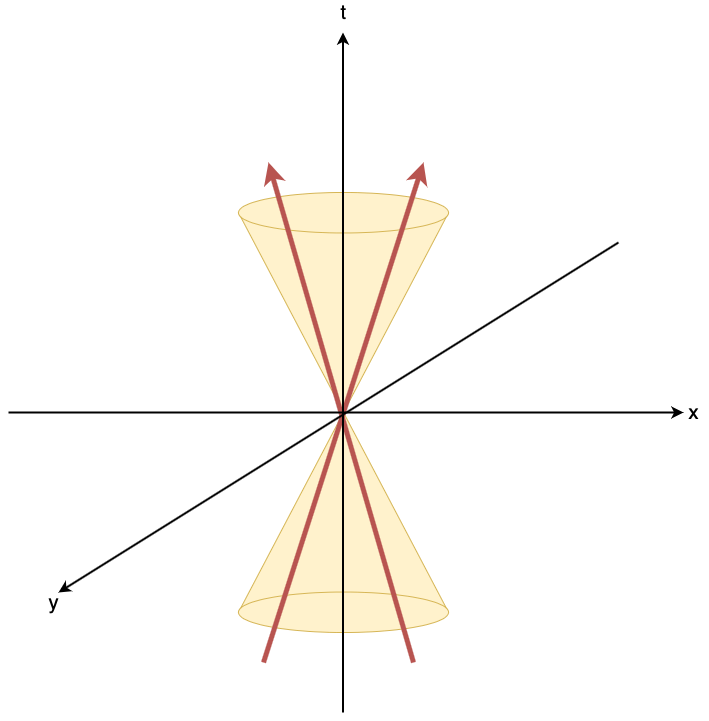}\label{lightcone_g} }}%
    \qquad
    \subfloat[\centering $\bar{g}_{\mu \nu}$]{{\includegraphics[width=6cm]{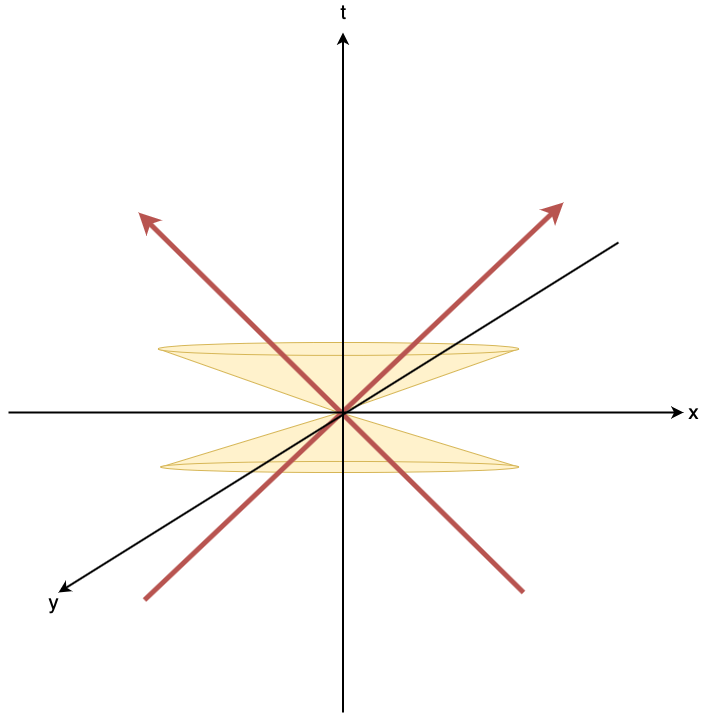}\label{lightcone_gbar} }}%
    \captionsetup{justification=raggedright}
    \caption{Light cone structures associated with different constant metric tensors. Time-like trajectories are represented by red arrows.}%
    \label{}%
\end{figure}

Before closing this section, we add  two comments; The first one is on DL cases using causal
structures of light cones which may appear in $\mathcal{I}_{M}$ for large $M
$ with a random choice of the geodesics. If we consider  time-space image data $\mathcal{I}_{M}^{(0,1)}$ of particle
trajectories in the $(x^{0},x^{1})$ plane with large $M$, it is possible for
the DL to notice the difference of the two metrics even if the corresponding $\Gamma
_{\nu\nu^{\prime}}^{\mu}\left(  x\right)  $'s coincide with each other.
For example, let us consider a Miknowski spacetime. Any constant Riemannian metric tensor $g_{\mu\nu}$ provides the same Christoffel symbols as $\Gamma^{\alpha}_{\beta\gamma}=0$. 
If we use  space-space image data $\mathcal{I}_{M}^{(1,2)}$ for DL, the two different constant tensors $g_{\mu\nu}\neq \eta_{\mu\nu}$ and $\bar{g}_{\mu\nu}=\eta_{\mu\nu}$ are connected by a DL isometry as seen Fig.~\ref{metric_g} and Fig.~\ref{metric_gbar}. Meanwhile, if a DL sees  time-space image data $\mathcal{I}_{M}^{(0,1)}$,  the two data have different light cone structures associated with $g_{\mu\nu}$ in Fig.~\ref{lightcone_g} and $\bar{g}_{\mu\nu}$ in Fig.~\ref{lightcone_gbar}. Thus, in this example, the DL is able to notice the metric difference. 
However, if $M$ is not so large, the DL often fails the discrimination since the light cone structure becomes blurred and cannot be seen clearly. If the $\lambda$ dependence appears only in a conformal factor of the metric, which has the maximal DL isometry like Eq.\eqref{e200}, 
any DL with large $M$, which sees $\mathcal{I}_{M}^{(0,1)}$, cannot notice the $\lambda$ dependence because the causal structure does not have $\lambda$ dependence. Note again that, if we consider a  space-space image data $\mathcal{I}_{M}^{(1,2)}$ with large $M$ in the
$(x^{1},x^{2})$ plane, the DL  never succeeds in noticing the difference of two
metrics $g_{\mu\nu}(x)$ and $\bar{g}_{\mu\nu}(x)$ as long as $\Gamma_{\nu
\nu^{\prime}}^{\mu}\left(  x\right)  =\bar{\Gamma}_{\nu\nu^{\prime}}^{\mu
}\left(  x\right) $.

 The second comment is on quantum aspect of the DL isometry. In quantum gravity, it may be considered that two spacetimes connected by a DL isometry are in quantum superposition. When DL metric detectors observe the superposed states, the post-measurement states may be still in the same superposition if a DL isometry exists. For example, let us consider two metrics $g_{\mu\nu}$ and $\bar{g}_{\mu\nu}$, and assume that the two metrics are connected by a DL isometry. Let us also assume that two quantum states $|g\rangle$ and $|\bar{g}\rangle$ correspond to $g_{\mu\nu}$ and $\bar{g}_{\mu\nu}$. Moreover, let us regard image data of geodesic trajectories of test particles as quantum states of a quantum system for the metric detection. The initial state of the image system is denoted by $|0\rangle$, and evolves into a quantum system $|\mathcal{I}_{g}\rangle$ for $g_{\mu\nu}$, and $|\mathcal{I}_{\bar{g}}\rangle$ for $\bar{g}_{\mu\nu}$. When a superposed quantum state $a|g\rangle+b|\bar{g}\rangle$ is observed by the DL metric detector, the composite system evolves as  
\begin{equation}
    \left(a|g\rangle+b|\bar{g}\rangle \right)\otimes|0\rangle \rightarrow a|g\rangle\otimes|\mathcal{I}_{g}\rangle+b|\bar{g}\rangle\otimes|\mathcal{I}_{\bar{g}}\rangle.
\end{equation}
If $|\mathcal{I}_{g}\rangle=|\mathcal{I}_{\bar{g}}\rangle$ holds owing to the DL isometry, the final state is given by$\left(a|g\rangle+b|\bar{g}\rangle\right)\otimes|\mathcal{I}_{g}\rangle$.
Hence the superposition of the state is preserved. On the other hand if $|\mathcal{I}_{g}\rangle\perp |\mathcal{I}_{\bar{g}}\rangle$ and DL detector discriminates $|\mathcal{I}_{g}\rangle$ and $ |\mathcal{I}_{\bar{g}}\rangle$, the total state is entangled and the post-measurement state collapses to $|g\rangle$ or $|\bar g\rangle$ depending on the measurement result. Hence the quantum superposition is destroyed.  
\section{DL Estimation of Cosmological Constant in Anti-de Sitter Spacetime}

In this section, we demonstrate a DL estimation of metrics of the AdS spacetime in
2+1 dimensions. The spacetime is crucial in the AdS/CFT context
\cite{ADSCFT}. 
Let us consider an AdS metric form corresponding to Eq.~\eqref{e2} as%
\begin{equation}
ds^{2}=g_{\mu\nu}(x,\lambda)dx^{\mu}dx^{\nu}=-\left(  1+\lambda^{2}%
r^{2}\right)  dt^{2}+\frac{dr^{2}}{1+\lambda^{2}r^{2}}+r^{2}d\phi^{2},
\label{e201}
\end{equation}
where $\lambda$ is a positive parameter, and connected with scalar curvature as $
R=-6\lambda^{2}$. Thus $\lambda$ indicates the square root of the absolute value of the negative cosmological constant in the AdS spacetime.

First let us prove that a nontrivial CCST $\delta g_{\beta\gamma}$ does not
exist in the spacetime using Eq.~\eqref{11}.  Here we do not need to assume that the spacetime dimension is three, and the proof is the same in any higher dimensions. Thus let us consider $D$
dimensional AdS spacetimes here. In the de Sitter spacetime case with a positive cosmological constant, the same
conclusion is obtained by replacing $\lambda^{2}\rightarrow-\lambda^{2}$. Since the spacetimes are maximally symmetric, the following relation is satisfied for the Riemannian curvature tensor:%
\begin{equation}
R_{\alpha\beta\mu\nu}=\frac{R}{D\left(  D-1\right)  }\left(  g_{\alpha\mu
}g_{\beta\nu}-g_{\beta\mu}g_{\alpha\nu}\right)  ,\label{20}%
\end{equation}
where $R$ is the scalar curvature. Substituting Eq.~\eqref{20} into
Eq.~\eqref{11} yields
$$
g_{\beta\mu}\delta g_{\gamma\nu}-g_{\beta\nu}\delta g_{\gamma\mu}+g_{\gamma
\mu}\delta g_{\beta\nu}-g_{\gamma\nu}\delta g_{\beta\mu}=0.
$$
By taking a trace in the above equation, we obtain
$$
\delta g_{\mu\nu}=\frac{1}{D}g_{\mu\nu}\left(  g^{\alpha\beta}\delta
g_{\alpha\beta}\right)  .
$$
Since $\delta g_{\alpha\beta}$ is assumed to be a CCST, $g^{\alpha\beta}\delta
g_{\alpha\beta}$ is a constant $\delta b$ independent of $x$. Thus the CCST
should be trivial:
\begin{eqnarray}
\delta g_{\mu\nu}=\frac{\delta b}{D}g_{\mu\nu}. \label{e300}
\end{eqnarray}
 Therefore, we do not need to worry about the DL isometry for the estimation of $\lambda$ except the trivial case in  Eq.~\eqref{e300}.
 
Note that if we consider a $\lambda$-dependent coordinate transformation given by $\tau =\lambda\, t$ and $\sigma= \lambda\, r$, the 2+1 dimensional AdS  metric form is given by
$$
ds^{2}=\frac{1}{\lambda^2} \left(-\left(  1+\sigma^{2}\right)  d\tau^{2}+\frac{d\sigma^{2}}{1+\sigma^{2}}+\sigma^{2}d\phi^{2}
\right).
$$
The above metric form depending on $\lambda$ has the maximal DL isometry satisfing Eq.~\eqref{e300}. But in the metric of Eq.~\eqref{e201}, no DL isometry exists for the estimation of $\lambda$ because $g^{\mu\nu}\delta g_{\mu\nu}$ has nontrivial dependence on $r$.

Next let us consider a DL for the $\lambda$ estimation in detail.  As seen in the Appendix B, the solution of the equation of motion of a free particle in the spacetime is given by
\begin{eqnarray}
r &=& \frac{1}{\lambda} \sqrt{ E - \sqrt{E^2-\lambda^2 L^2} \sin (2\lambda \tau)} \label{eq:r_sol}, \\
\phi &=& \pm  \arctan \left[\frac{E}{\lambda L} \left(\tan (\lambda \tau) - \sqrt{1 - \left(\frac{\lambda L}{E}\right)^2} \right)  \right] + \phi_0 \label{eq:phi_sol}, \\
t &=& \frac{1}{\lambda} \arctan \left[  \frac{(1+E)\tan (\lambda \tau) -  \sqrt{E^2-\lambda^2 L^2} }{\sqrt{1+2E + \lambda^2 L^2}}\right]+ t_0 \label{eq:t_sol},    
\end{eqnarray}
where $E$ is a positive constant associated with energy, $L$ is a real constant associated with angular momentum, and $t_0$ ($\phi_0$) is an integration constant for $t$ ($\phi$ ). 

\begin{figure}[H]
	\centering
	\includegraphics[width=7cm]{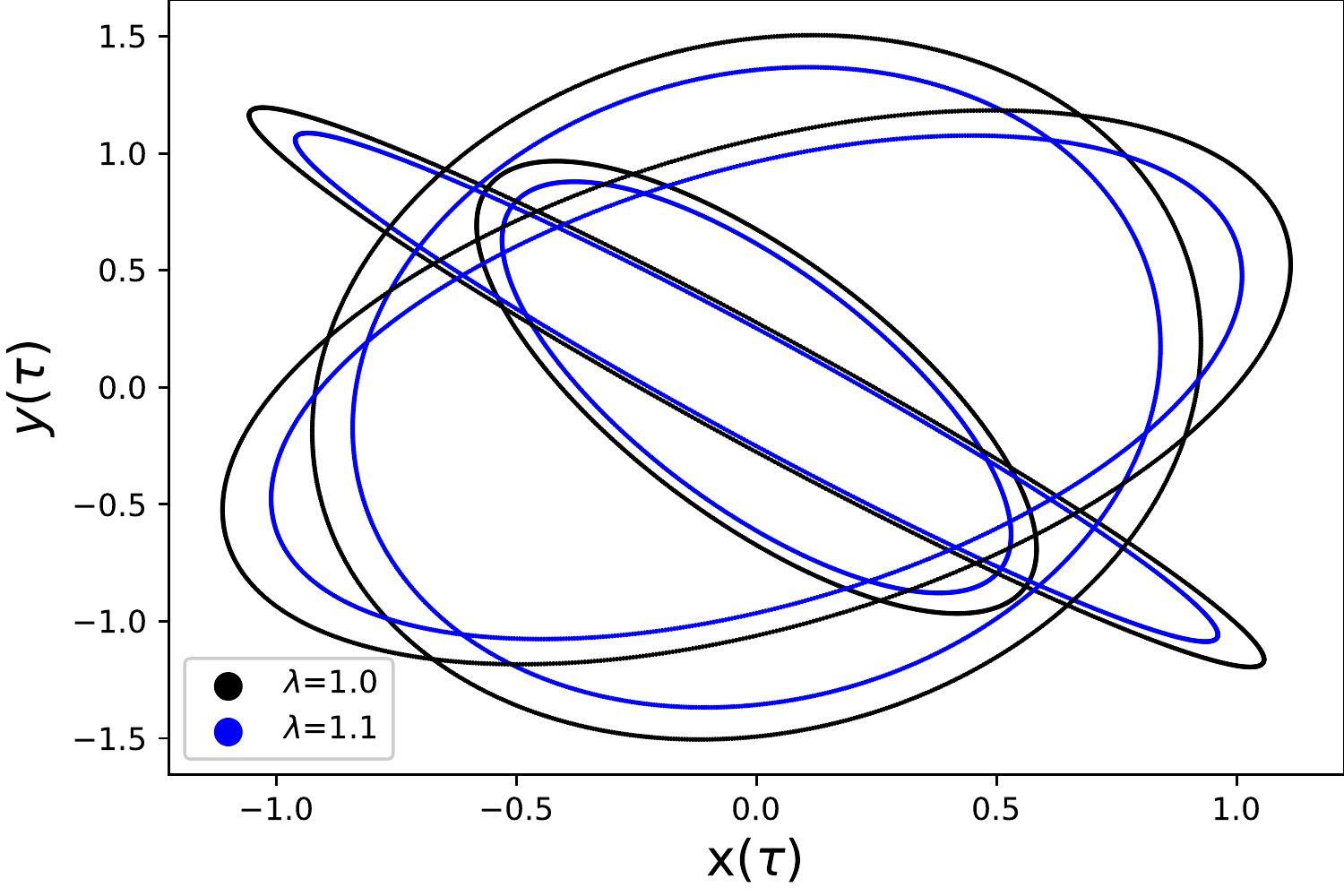}
	\caption{Two images of $\mathcal{I}_{M}$ with $\lambda= 1$ (black line) and $\lambda=1.1$ (blue line)  with $M= 4$ (number of geodesics).}
	\label{geodesics}
\end{figure}

For each fixed value of $\lambda$, let us generate an image data $\mathcal{I}_{M}$
of $M$ geodesics in a Cartesian coordinate system spanned by  $\left(
x, y\right) =\left(
r \cos{\phi},r \sin{\phi}\right)  $. For instance, in Fig.~\ref{geodesics}, the geodesics for two different  $\lambda$ values,  $\lambda=1$ and  $\lambda=1.1$, are plotted for $M=4$ (number of geodesics). According to Eq.~(\ref{eq:r_sol}), the length of the semi-major axis of the trajectory ellipse decreases as $\lambda$ increases. In other words, the elliptical form of the geodesic trajectory shrinks in a larger $\lambda$ spacetime. On the other hand, if we consider a flat spacetime with $\lambda=0$, the particle runs along a straight line. The data is treated as a training data of our supervised DL.
After the training, we input a new $\mathcal{I}_{M}$ to the  DL, and the output of the estimated value for $\lambda$ is obtained.

Here, we explain the method we used in the simulation. The procedure is as follows. First we generate a data set using the analytical solution of the equation of motion. Then we split the data  into training data, validation data and test data. Training data are used for training our DL model. Validation data are utilized to check whether the DL model overfits  the training part. The test data are used to evaluate the model performance. 

As mentioned in section 1, the mechanism of DL being capable to predict the cosmological constant $\lambda$ under any local coordinate transformations is exactly equivalent to that of recognizing handwritten digits. Let us imagine the digits ``2" in Fig.~\ref{fig:MNIST}, for instance, are actually under 9 different local transformations. DL is still going to give the correct digit regardless of any local transformations. In a general sense, DL learns huge numbers of examples with different writing styles to infer rules for recognizing handwritten digits. Following the same logic, it is promising that DL is able to give the correct cosmological constant (digit) under any local coordinate transformations (writing styles).

\begin{figure}[]
	\centering
	\includegraphics[width=12cm]{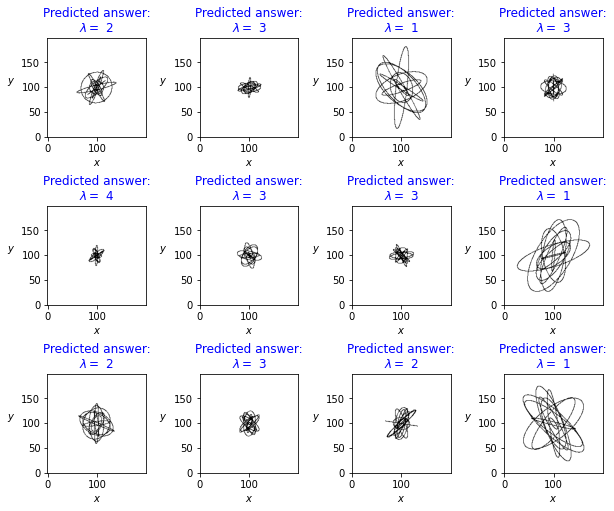}
	\caption{Example of images of trajectories in the $(x,y)$ plane of in test data and predictions from DL.
	The horizontal axis is $x$, and the vertical axis is $y$. Each axis is discretized by 200 pixels in the range $x_{\text{min}}\le x \le x_{\max}$ and $y_{\min} \le y\le y_{\max}$, where $x_{\min}$ and $y_{\min}$ are the minimum values of $x$ and $y$, respectively, and $x_{\max}$ and $y_{\max}$ are the maximum values of $x$ and $y$, respectively.
The title of each image shows the prediction and the blue (red) color of the title means the correct (incorrect) prediction. DL returns correct answers for presented images. \hfill}%
\label{fig:data_used}
\end{figure}

\begin{figure}[htb]
	\centering
	    \captionsetup{justification=raggedright}
	\includegraphics[width=12cm]{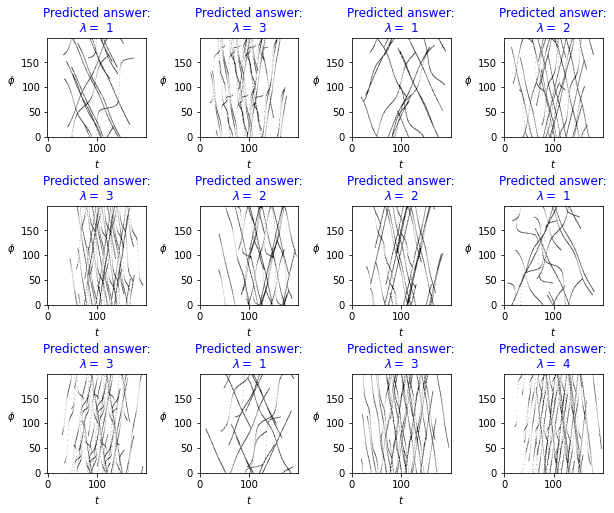}
	\caption{Example of images of trajectories in the $(t, \phi)$ plane of  test data and predictions from DL. Each axis is discretized by 200 pixel in the range $ t_{\min} \le t \le t_{\max} + 0.01$ , $ 0 \le \phi\le 2\pi $, where the $ t_{\min} $ and $ t_{\max}$ are the minimum value and the maximum value of $t$, respectively.  The title of each image shows the prediction. If the predication coincides with the model answer, the color of the title is in blue, otherwise it is in red. The DL returns only correct answers for presented images.}
	\label{fig:time-phi}
\end{figure}

In the following, we show our results of simulations. The scientific software needed to reproduce the results presented below may be found in an open source GitHub repository \cite{github}. First, we prepare 2000 image data, each data  consists of 200 $\times$ 200 pixels. We set $\lambda$ to be discretized and an integer between 1 and 4 for simplicity in the simulation. The value for $\lambda$ of each image is determined by a uniformly distributed random number. By taking $\tau_0 = 0$, a particle trajectory is determined by giving two conserved quantities $L$ and $E$, and the value of $\phi$ at $\tau = \tau_0$. The values of $L$, $E$ and $\phi_0$ take uniformly distributed random numbers. The ranges of these  number are $0\leq E < 10$, $0 \leq L < E/\lambda $ and $-\pi \leq \phi_0 < \pi$, respectively. We split whole data into three parts: 1400 training data, 400 validation data and 200 test data. Then we train a CNN model using the data with 100 epoches. The loss at the end declines to 0.0012 and the accuracy of the trained model for test data is about 96\%. The first 12 images in the test image data of 10 particle trajectories are depicted in Fig.~\ref{fig:data_used}. For every image, the value of $\lambda$ with the highest possibility is chosen and shown in the title. If the predication coincides with the expected value, the color of the title will be in blue, otherwise it will be in red. 
\noindent
We also check whether or not the DL can predict $\lambda$ using $(t, \phi)$ images instead of $(x,y)$ images. We generated 2800 training data, 800 validation data and 400 test data. The value of $\lambda$ is chosen as an integer between 1 and 4. In this case, the accuracy for the test data is also high and reaches almost 100\%. The first 12 test data are illustrated in Fig.~\ref{fig:time-phi}.

\begin{figure}[htb]
\centering
    \captionsetup{justification=raggedright}
\includegraphics[width=7cm]{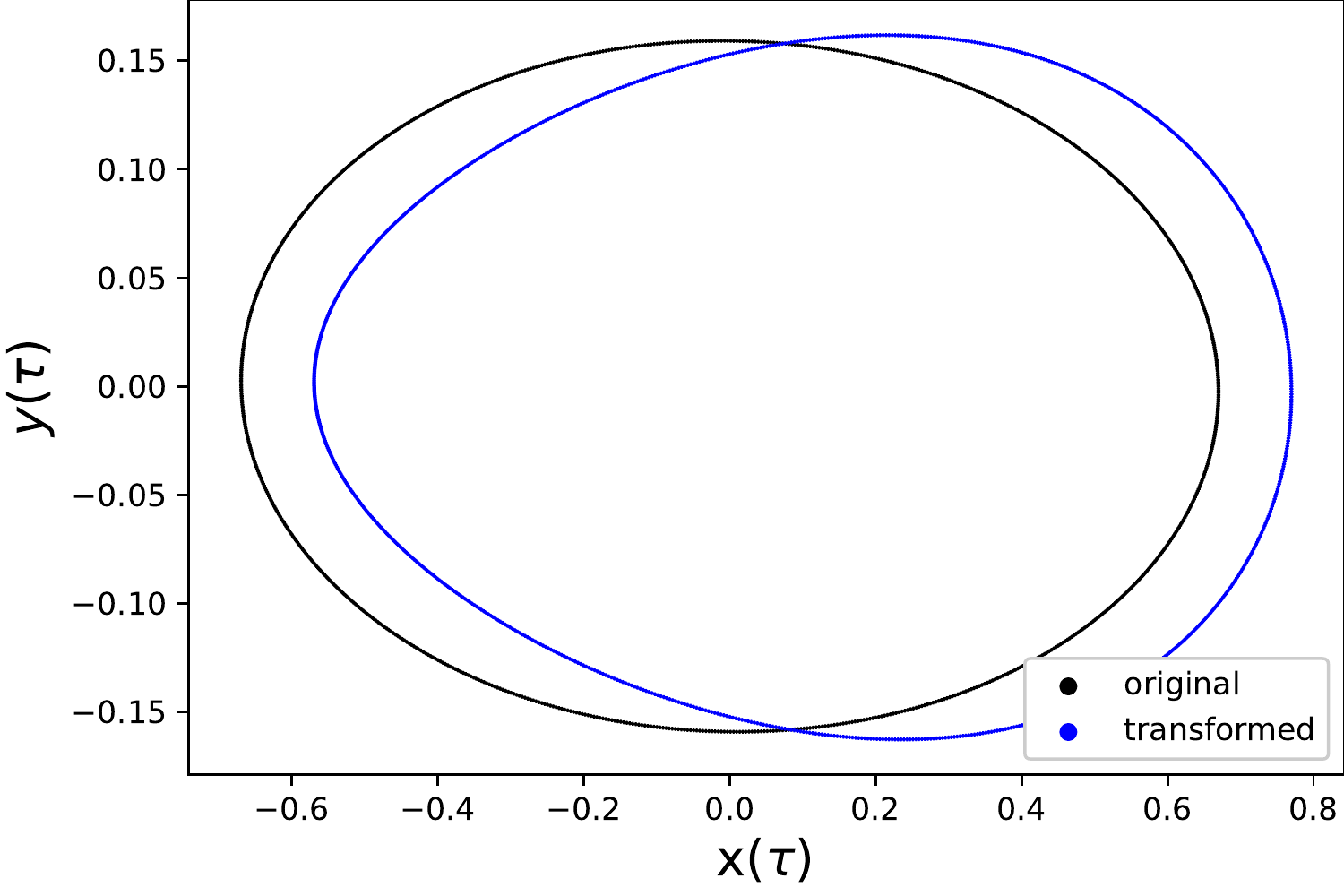}
	\caption{Particle trajectories connected by a local coordinate transformation with $x'=x+0.1\cos(nxy)$ and $y'=y+0.1\sin(nxy)$, where $n=3$. The black trajectory is the original one, and the blue trajectory is the transformed one.}
	\label{local_transf}
\end{figure}

\begin{figure}[htb]
	\centering
	    \captionsetup{justification=raggedright}
	\includegraphics[width=12cm]{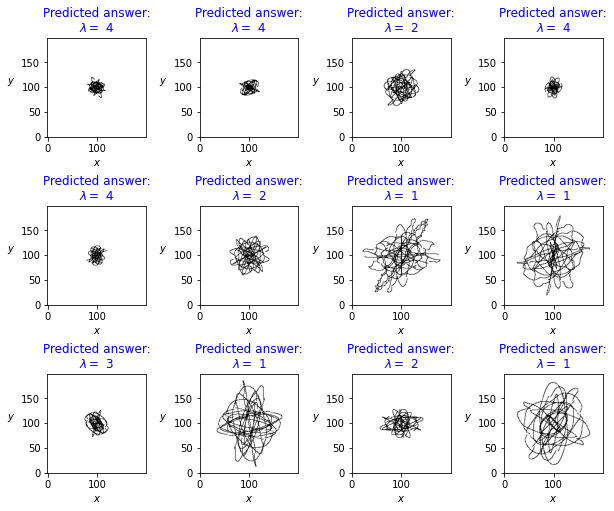}
	\caption{ Examples of images of trajectories in the $(x, y)$ plane of  test data which are locally  transformed. The DL can estimate correct values of $\lambda$ even though the local transformation is included. }
	\label{fig:local_result}
\end{figure}

It is also verified that the DL metric detector is not sensitive to the freedom of local coordinate transformations and predicts the same value of $\lambda$ for two different image data like in Fig.~\ref{local_transf}. In the simulation, we consider the following local coordinate transformation:
\begin{eqnarray}
x^{\prime} &=& x + 0.1 \cos(nxy) ,  \notag \\
y^{\prime} &=& y + 0.1 \sin (nxy), \label{eq:local_trans}
\end{eqnarray}
where $n$ is an integer between 1 and 4.
We first generated non-transformed 4000 data. Again, we set $\lambda$ to an integer between 1 and 4 as in the above simulation. Then, we applied the local coordinate transformation given by Eq. (\ref{eq:local_trans}), where $n$ was selected at random from an integer between 1 and 4, to each image.  The data contained four types of locally transformed data.  We split the data into the training data,  validation data and  test data. The number of those were 2800, 800 and 400, respectively.
The trained DL model returned correct answers with about 99\% accuracy for the test data. Examples of the test data and the prediction values of the DL are illustrated in Fig. \ref{fig:local_result}.
The demonstration shows that the DL succeeds in neglecting the local coordinate transformation.

\section{DL Estimation of Asymptotic Symmetry Charges in AdS}

In the AdS/CFT correspondence, the 2+1 dimensional pure gravity theory is dual to a 1+1 dimensional CFT which lives at spatial infinity and possesses a positive central charge $c$. In the large $c$ limit, the gravity theory approaches the classical GR theory. The asymptotic isometry of the bulk spacetimes consists of two sets of Virasoro symmetries \cite{BH}. The symmetries are referred to as Brown-Henneaux (BH) symmetry, and the associated physical charges are denoted by $Q$ in this paper. Different spacetimes described by the asymptotic metrics can be discriminated by $Q$. The CFT quantum states corresponding to  different spacetimes are also discriminated by their corresponding BH charges $Q_{CFT}$ in CFT.  On the CFT side, it is possible to adopt the standard quantum measurement theory, which includes the concepts of measurement operators and positive operator valued measure (POVM) \cite{NC}. Some of the CFT quantum measurements are capable of detecting $Q_{CFT}$. The values of $Q_{CFT}$ enable us to reproduce the bulk metrics with the same values of $Q$ \cite{QM}. Thus a part of the CFT quantum measurement plays the role of a metric detector. This implies that the CFT quantum measurement theory can be interpreted as a quantum measurement theory for the AdS quantum gravity.
Then a natural question arises: What kinds of DL metric detecrors in the bulk correspond to CFT quantum measurement devices for $Q_{CFT}$? The corresponding measurement of the CFT may be a quantum DL measurement. Then another question is following: Is any DL measurement device for $Q_{CFT}$ dual to a bulk DL metric detector?  No complete answers for these questions exist at present. But we are able to provide interesting speculations about that.  Note first that the values of $Q$ are completely evaluated only by the near-boundary information in the bulk. Thus it might be tempting to consider that bulk DL's, which explore only the near-boundary region, correspond to the CFT DL's.   Contrary to this naive  expectation, this may not be the case. Two bulk metrics connected by the asymptotic isometry are almost the same near the spatial infinity boundary. This reminds us of the argument of the DL isometry in section 2. In the near-boundary region, the asymptotic isometry behaves as an approximate DL isometry and causes a serious difficulty in the metric discrimination. Since the deviation between the two metric forms is very small near the boundary, the geodesic image data merely show a very tiny difference near the boundary. Thus, only by using the near-boundary data, the estimation of $Q$ is a quite hard task for any DL and loses its efficiency.  

On the other hand, suppose that for CFT some DL's exist which discriminate the $Q_{CFT}$ valules with high efficiency. No obstacle appears to assume the existence of such high-efficiency DL's for the CFT since the conformal fields are merely ordinary quantum systems, and $Q_{CFT}$ is also one of the ordinary observables. Then one may expect that the quantum DL for CFT are capable of clearly discriminating bulk metrics connected by the asymptotic isometry. This means that the corresponding metric detectors should utilize  not only the near-boundary information but also the deep-region information. Thus, at least in the classical GR regime with large $c$, the DL metric detectors require metric forms $g_{\mu\nu}(x,Q)$ which are parametrized by $Q$ and vary drastically in the deep-region far from the boundary as the values of $Q$ change. In this sense, a gauge fixing becomes significant in the region to realize such a specific form of $g_{\mu\nu}(x,Q)$. One might think that this is strange since the metric deviation in the deep region should be described by a local coordinate transformation, i.e., unphysical gauge transformation. However, the above speculation about DL suggests that the AdS/CFT correspondence should treat such a gauge freedom in the bulk region as an actual physical freedom. This is a quite nontrivial indication for the AdS quantum gravity.
The similar claim was emphasized in \cite{QM} based on a different argument from our DL measurement argument. This deep-region feature of the AdS space may become more understandable in the context of gravitational dressing \cite{QM2}. About the second question, coherent states for Virasoro orbits \cite{Nair} may be related to the answer. The coherent states play a role of an overcomplete basis in the Verma module state space of  CFT. Thus it is possible to construct measurement operators by using the projection operators of the coherent states. When the central charge $c$ is large, the CFT coherent states discriminated by the quantum expectation value $\langle Q_{CFT} \rangle$ of $Q_{CFT}$ may correspond to classical bulk metrics with $Q=\langle Q_{CFT} \rangle$. 

In the later part of this section, we explicitly demonstrate a DL metric detection for $Q$ in the bulk. The asymptotic AdS metric $ds^{2}=g_{\mu\nu}dx^{\mu}dx^{\nu}$ is described by
the fall-off conditions for metric components in \cite{BH}. When a polar coordinate system
spanned by $\left(  t,r,\phi\right)  $ is adopted, the fall-off condition is
given by%
$$
\left[  g_{\mu\nu}\right]  =\left[
\begin{array}
[c]{ccc}%
g_{tt} & g_{tr} & g_{t\phi}\\
g_{rt} & g_{rr} & g_{r\phi}\\
g_{\phi t} & g_{\phi r} & g_{\phi\phi}%
\end{array}
\right]  =\left[
\begin{array}
[c]{ccc}%
-\lambda^{2}r^{2}+O(r^{0}) & O(r^{-3}) & O(r^{0})\\
O(r^{-3}) & \frac{1}{\lambda^{2}r^{2}}+O(r^{-4}) & O(r^{-3})\\
O(r^{0}) & O(r^{-3}) & r^{2}+O(r^{0})
\end{array}
\right]  ,
$$
where $r=\infty\,~$represents the spatial infinity boundary and $\phi$ is
angular variable. The point satisfying $\phi=2\pi~$is identified as the point
satifying $\phi=0$. In this coordinate, the exact AdS metric is given by Eq.~\eqref{e201}. Let us consider the following vector field $\xi$ as a generator of the BH  asymptotic isometry \cite{BH}:
\begin{eqnarray}
\xi^{(3)t} &=& \frac{1}{\lambda} \xi (\lambda t -\phi) +\frac{1}{2\lambda^3 r^2} \xi''(\lambda t -\phi)+O(r^{-4}), \label{01}\\
\xi^{(3)r} &=& -r \xi'(\lambda t -\phi)+O(r^{-1}),\label{02}\\
\xi^{(3)\phi} &=& - \xi (\lambda t -\phi) +\frac{1}{2\lambda^3 r^2} \xi''(\lambda t -\phi)+O(r^{-4}) \label{03},
\end{eqnarray}
where $\xi (\sigma)$ is a real function of $\sigma$ satisfying $\xi (\sigma+2\pi)=\xi (\sigma)$.
By using the ADM variables $N$, $N^a$ and $h_{ab}$ of the metric as
\begin{eqnarray}
ds^2 = -N^2 dt^2 +h_{ab}(dx^a +N^a dt)(dx^b+N^b dt),
\end{eqnarray}
$\xi$ is also described by other vector fields as
\begin{eqnarray}
\xi^{\perp} &=& N \xi^{(3)t},\\
\xi^{r} &=& \xi^{(3)r}+N^r \xi^{(3)t},\\
\xi^{\phi} &=& \xi^{(3) \phi}+N^\phi \xi^{(3)t}.
\end{eqnarray}
Adopting the standard ADM formalism, we are able to define the BH charge $Q$ for the asymptotic metrics as follows:
\begin{eqnarray}
Q[\xi]=\lim_{r\rightarrow \infty}\oint dS_l 
\left( \bar{G}^{ijkl}\left(\xi^\perp \nabla^{(3)}_{k} g_{ij} -\partial_k \xi^\perp (g_{ij} -\bar{g}_{ij}) \right)
+2\xi^i \Pi_{i}^l
\right),
\end{eqnarray}
where $\bar{G}^{ijkl}$ is the ADM superspace metric tensor for the spatial section metric $h_{ab}$, $\Pi_{i}^l$ is the conjugate momentum tensor, and $\bar{g}_{ij}$ are the spatial components of the exact AdS metric as
\begin{eqnarray}
ds^2 =\bar{g}_{\mu\nu} d\bar{x}^\mu d\bar{x}^\nu = -(1+\lambda^2 \bar{r}^2) d\bar{t}^2 
+ \frac{d\bar{r}^2}{1+\lambda^2 \bar{r}^2} + \bar{r}^2 d\bar{\phi}^2.\label{-1}
\end{eqnarray}
By repeating the infinitesimal coordinate transformation generated by the vector field $\xi$ in Eqs.~(\ref{01})-(\ref{03}),  finite transformations are generated. For instance, one of the transformations is computed by using a  monotocially increasing function $F(\sigma)$ satisfying the following conditions:
\begin{align*}
    F(\sigma +2\pi)=F(\sigma)+2\pi,\quad
    F'(\sigma +2\pi)=F'(\sigma),\quad
    F''(\sigma +2\pi)=F''(\sigma),
\end{align*}
where the prime (double prime) mean the first (second) derivative with respect to $\sigma$.
Then it turns out that the following is a finite coordinate transformation of the BH asymptotic isometry:
\begin{align}
&\bar{t} = \frac{1}{2\lambda}(\lambda t +\phi)+\frac{1}{2\lambda}F(\lambda t -\phi)
+\frac{1}{4\lambda^3 r^2 } F''(\lambda t -\phi),\label{04}\\
&\bar{r} = \frac{r}{\sqrt{F'(\lambda t -\phi)}},\label{05}\\
&\bar{\phi} = \frac{1}{2}(\lambda t +\phi)-\frac{1}{2}F(\lambda t -\phi)
+\frac{1}{4\lambda^2 r^2 } F''(\lambda t -\phi).\label{06}
\end{align}
Note that Eqs.~(\ref{01})-(\ref{03}) are recovered by taking $F(\sigma)=\sigma +2\xi(\sigma)$ in the above equations and computing the linear contributions $O(\xi)$. We get asymptotic metrics $g_{\mu\nu}(x,Q)$ by applying the transformation to the exact AdS metric in Eq. (\ref{-1}). The value of $Q$ is directly computed for each vector field $\xi$. 
When we take
$$
\xi(\sigma)=\xi_{n+} (\sigma)=\cos (n\sigma)
$$
in Eqs. (\ref{01})-(\ref{03}), the charge is computed as 
\begin{equation}
    Q[\xi_{n+}]
    =\frac{2}{\lambda} \int^{2\pi}_{0} d\sigma \cos(n\sigma)\left[\partial_{\sigma}\left(\frac{F''(\sigma)}{F'(\sigma)}\right)-\frac{1}{2}\left(\frac{F''(\sigma)}{F'(\sigma)}\right)^2\right]. \label{c1}
\end{equation}
When we take
$$
\xi(\sigma)=\xi_{n-} (\sigma)=\sin (n\sigma)
$$
in Eqs. (\ref{01})-(\ref{03}),  the charge is computed as 
\begin{equation}
    Q[\xi_{n-}]
    =\frac{2}{\lambda} \int^{2\pi}_{0} d\sigma \sin (n\sigma)\left[\partial_{\sigma}\left(\frac{F''(\sigma)}{F'(\sigma)}\right)-\frac{1}{2}\left(\frac{F''(\sigma)}{F'(\sigma)}\right)^2\right].
\end{equation}

\begin{figure}[htb]
\centering
\includegraphics[width=8cm]{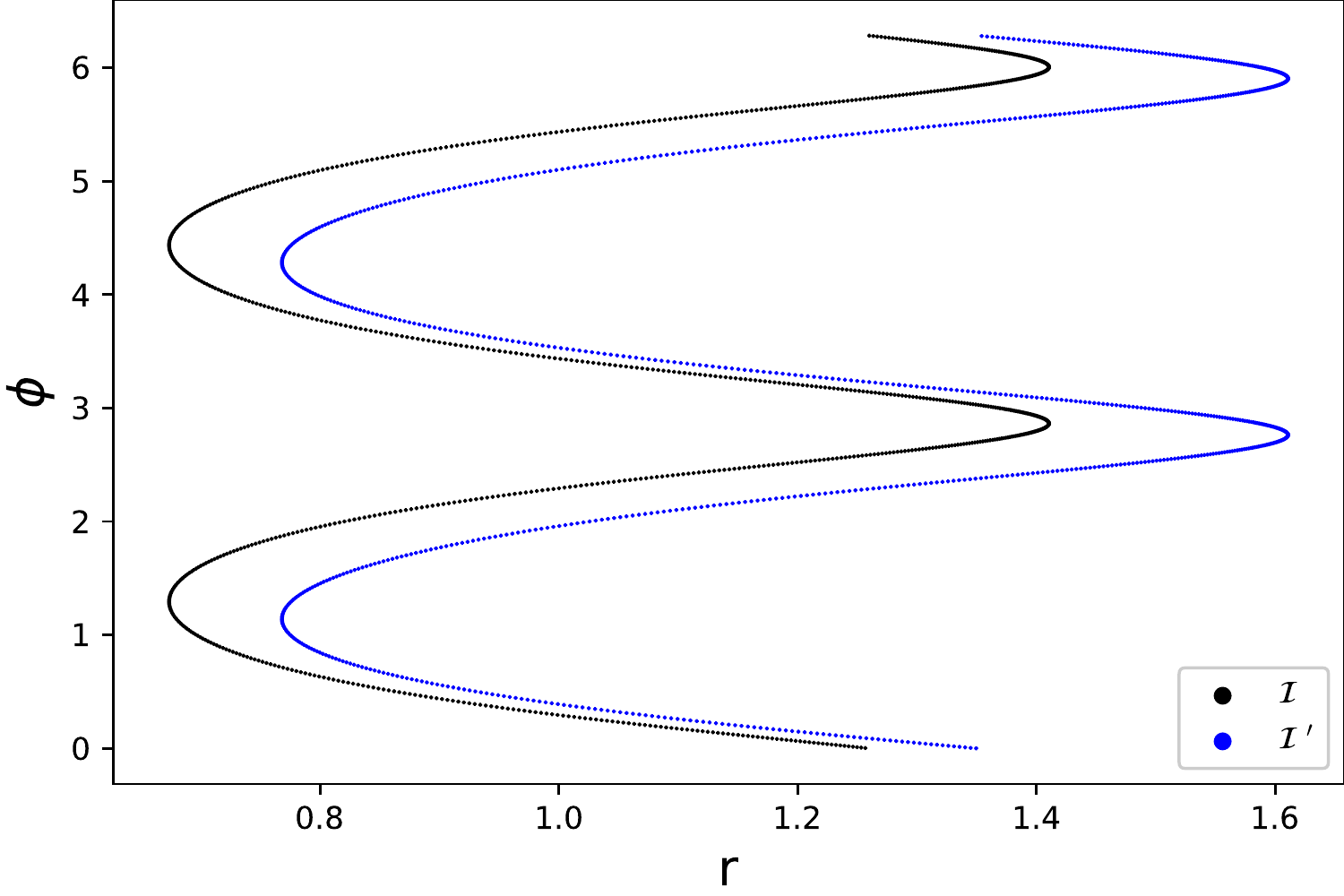}
	\caption{Particle trajectories connected by the coordinate transformation of the asymptotic isometry \eqref{04}-\eqref{06} when we take  $F = \left( \lambda_{1} t-\phi\right) + \lambda_{2} \sin \left( \lambda_{1} t-\phi\right)$ with $\lambda_{1}=1$  and $\lambda_{2}=0.3$. The black line is the original trajectory and the blue line represents the transformed one. }
	\label{BH_transf}
\end{figure}
\begin{figure}[htb]
	\centering
	\includegraphics[width=12cm]{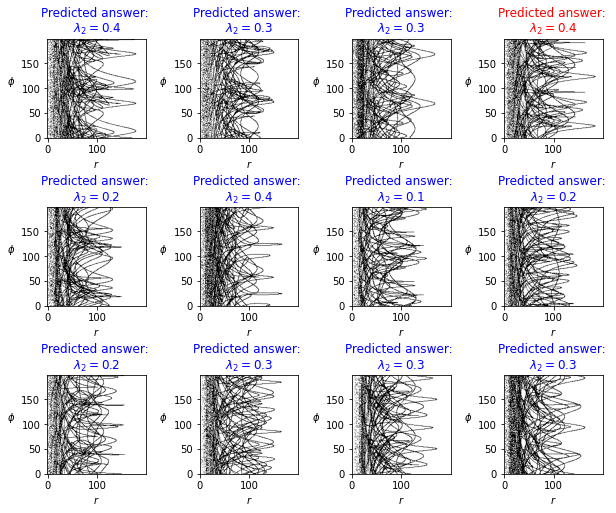}
	\caption{Examples of images of trajectories in the $(r, \phi)$ plane of test data which are locally transformed. The horizontal direction in each panel indicates the $r$ axis, and the vertical direction indicates the $\phi$ axis. Each axis is discretized by 200 pixels. The title of each image shows the prediction. If the predication coincides with the model answer, the color of the title is in blue, otherwise it is in red. In the above examples except for one image, the DL correctly answers the true values of $\lambda_2$, which appear on top of each panel. }
	\label{fig:bh_result}
\end{figure}

Later, the cosmological constant $\lambda$ is rewritten as $\lambda_{1}$. Let us focus on the BH charge in Eq. (\ref{c1}). We first generate an $M$-particles image data  $\mathcal{I}_{M}$ generated by the exact AdS metric. Next we generate another image data $\mathcal{I'}_{M}$ by using the transformation in Eqs. (\ref{04})-(\ref{06}) with $\lambda\equiv\lambda_1$. When we fix the explicit function form of $F(\sigma)$, it is possible to demonstrate the DL. For example, let us set the function $F(\sigma)$ as 
\begin{equation}
    F\left(\sigma \right) = \sigma+ \lambda_{2} \sin\sigma, \label{07}
\end{equation}
where $\lambda_2$ is another parameter of the metric tensor as mentioned in section 2. The relation between the BH charge and $\lambda_2$ is obtained by Eq. (\ref{c1}) and Eq. (\ref{07}). Hence, determining $\lambda_2$ is equivalent to determining the BH charge. The obtained trajectories are plotted in Fig.~\ref{BH_transf}, where the transformed trajectory is stretched and shifted. The difference between $\mathcal{I}_{M}$ and $\mathcal{I'}_{M}$ is , in principle,  obvious for DL to observe. In general, the image data $\mathcal{I'}_{M}$ is prepared in  the $\left(
r,\phi\right) $ coordinate system as it gives more accurate predictions than those in the $\left(x,y\right) $ coordinate system. Following the similar method in section 3, we first generated 4000 images. The transformation given in Eqs. (\ref{04})-(\ref{06}) with Eq. (\ref{07}) is then applied to the images, where $\lambda_1$ is fixed to be 1 and $\lambda_2$ is characterized by 0.1, 0.2, 0.3 and 0.4. The number of orbits in one image is 25 and the size of an image is 200 $\times$ 200. We also split the data into training data,  validation data and  test data. We train the DL to predict numbers which characterize the local transformation.  The result is that the accuracy for test data is about 85\%.  Examples of the test data and the prediction values of the DL are illustrated in Fig. \ref{fig:bh_result}.
Predicting the BH charge, therefore, is more difficult than other task we demonstrated, but our result verifies  that the DL has a large potential to estimate the BH charges \cite{github,comment}.


\section{Summary}

In this paper, we investigate whether deep learning (DL) is in principle able to discriminate two different metrics $g_{\mu\nu}(x)$ and $\bar{g}_{\mu\nu}(\bar{x})$  using image data of geodesics. We first propose a concept of the DL isometry in section 2.  If the trajectories of particles in two spacetimes are the same, but nevertheless $g_{\mu\nu}(x) \ne \bar{g}_{\mu\nu}(\bar{x})$, no DL model succeeds in discriminating two metrics, and  we say that $g_{\mu\nu}(x)$ and $\bar{g}_{\mu\nu}(\bar{x})$ are connected by the DL isometry. The necessary condition of the maximal DL isometry is given by Eq. (\ref{e11}). Next we parametrize the metric by $\lambda = (\lambda_1, \cdots, \lambda_N)$ as in Eq. (\ref{e2}).  Then the DL isometry condition is rewritten by Eq. (\ref{2}) considering an infinitesimal deviation of $\lambda$,   meaning that  $\delta g_{\alpha \beta} $ is a covariant constant symmetric tensor (CCST) associated with $g_{\alpha \beta}$, and $g_{\alpha \beta}(x,\lambda) $ and $g_{\alpha \beta}(x,\lambda) +\delta g_{\alpha \beta}(x,\lambda)  $ are connected by the maximal DL isometry. If the deviation $\delta g_{\alpha\beta}$ is generated by a some Lie transport
of $g_{\alpha\beta}$ associated with a vector field $\epsilon^{\mu}~$such that
$\delta g_{\alpha\beta}=\nabla_{\alpha}\epsilon_{\beta}+\nabla_{\beta}%
\epsilon_{\alpha}$, then Eq.~\eqref{2} is recasted into Eq. (\ref{3}). If $\epsilon^{\mu}$ is a homothetic vector which satisfies Eq. (\ref{6}),  we find that with Eq. (\ref{5}) $\epsilon^{\mu}$ is a conformal Killing vector. Three example of DL isometries are  shown in Eqs. (\ref{7}) - (\ref{e200}). We also provide the integration formula Eq. (\ref{12}) for $\delta g_{\alpha\beta}$ which satisfies Eq. (\ref{11}). 

In section 3, we demonstrate a DL estimation of metrics of the AdS spacetime in 2+1 dimensions. We consider the parametrized metric form given by Eq. (\ref{e201}). We study the property whether DL is able to predict the value of $\lambda$, which is associated with the scalar curvature.  We proved that a non-trivial CCST $\delta g_{\beta \gamma}$ does not exist in the AdS spacetime  by using Eq. (\ref{11}) and Eq. (\ref{20}). Therefore we do not need to take care of the DL isometry for the estimation of $\lambda$ in this case except the trivial case given by Eq. (\ref{e300}). We generate data of trajectories using the solution of the equation of motion of a free particle in the spacetime Eqs. (\ref{eq:r_sol})-(\ref{eq:t_sol}). We find that  DL is able to predict $\lambda$ with high probability like Fig. \ref{fig:data_used} and Fig. \ref{fig:time-phi} using $(x,y)$ plane images and $(t,\phi)$ plane images of trajectories, respectively. DL also estimates $\lambda$ precisely if a local coordinate transformation is applied to images of trajectories presented in Fig. \ref{fig:local_result}.

In section 4, we demonstrate the application of DL for the estimation of 2+1 dimensional AdS asympototic symmetry charges $Q$, which are called Brown-Henneaux (BH) charges.  Applying the AdS/CFT correspondence, we are able to discriminate two asymptotic metrics with different values of $Q$ by measuring the corresponding observables $Q_{CFT}$ in the CFT side.  The two metrics connected by the asymptotic isometry are almost the same near the spatial infinity boundary. This causes an approximate DL isometry in the near-boundary region. Thus the estimation of $Q$ becomes hard for the DL metric detectors  by use of the near-boundary condition only. This means that  high-efficient metric detectors should utilize not only the near-boundary information but also the AdS deep-region information. The detectors require a gauge fixing such that the form of $g_{\mu\nu}$ in the deep region possesses sensitive dependence of $Q$.  Such DL detectors actually succeed in estimating the value of $Q$ as verified in the demonstration in Fig. \ref{fig:bh_result}.  This result may help us to develop the understanding of quantum measurements of $Q_{CFT}$ in the CFT at the AdS boundary.

\acknowledgements{
We would like to thank T. Houri and T. Takayanagi for useful discussions. We appreciate warm hospitality of YITP at the first extreme universe school (held from 3rd until 5th March 2022), where we discussed the subject of this paper. This research was partially supported by JSPS KAKENHI Grants, No. JP19K03838 (M.H.), 21H05188 (M.H.), 19K03866 (Y.N.), Foundational Questions Institute (M.H.),  Silicon Valley Community Foundation (M.H.), JST SPRING, Grant Number JPMJSP2114 (R.K.), a Scholarship of Tohoku University, Division for
Interdisciplinary Advanced Research and Education (R.K.), and the WISE Program for AI Electronics, Tohoku University (R.K.).}


\section*{Appendix A}
In this appendix, we  briefly review deep learning (DL). First, we review convolutional neural networks (CNN) which we will use in numerical analyses. After that, we describe the concept and actual flow of DL. More detailed description is included in standard textbooks for DL \cite{Bishop, Goodfellow}.

\subsection{Review of CNN}
The CNN is frequently used in image recognition \cite{Fukushima,LeCun,AlexNet,VGGNet}. The CNN is comprised of neural networks (NN),  convolution layer, and pooling layer.  First, we explain the convolution operation.  

\begin{figure}[htb]
 \centering
   \includegraphics[width=100mm]{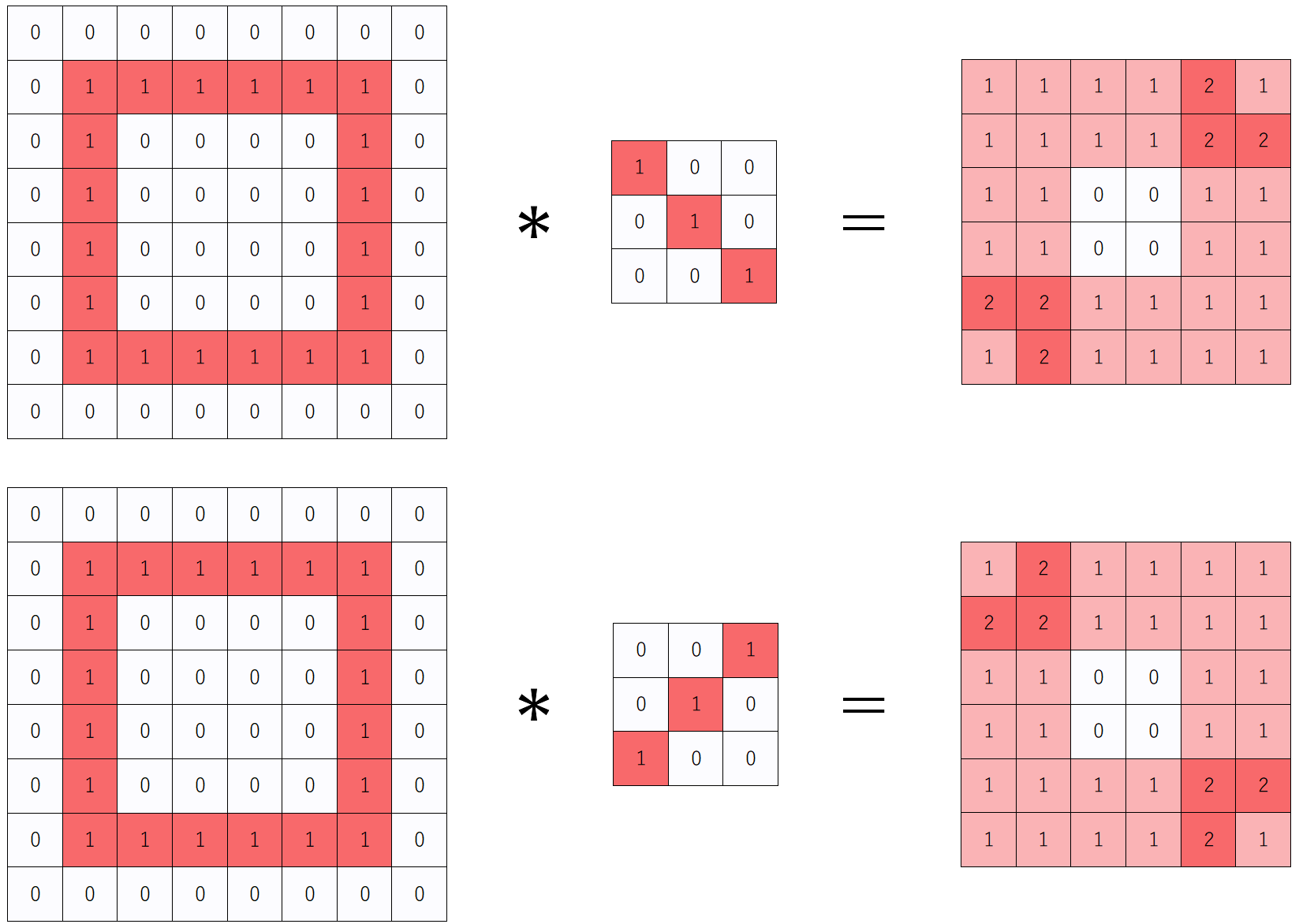}
 \caption{The example of convolution operations for $\mathcal{A}$ including a square and two convolution filters $\mathcal{C}$.}
 \label{fig:conv1}
\end{figure}

\begin{figure}[htb]
 \centering
   \includegraphics[width=100mm]{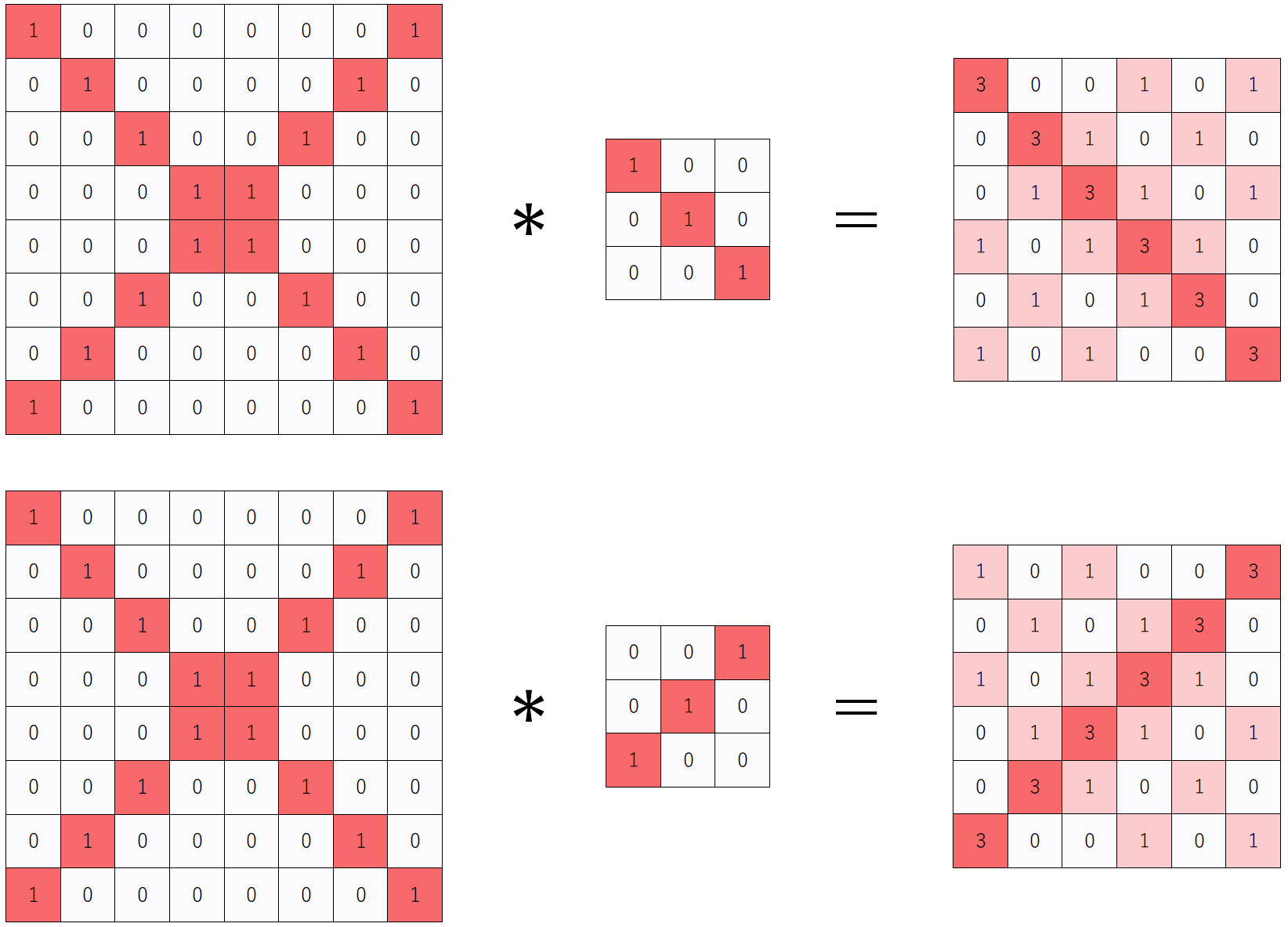}
 \caption{The example of convolution operations for $\mathcal{A}$ including a cross and two convolution filters $\mathcal{C}$.}
 \label{fig:conv2}
\end{figure}

Let us consider a two-dimensional matrix given by
\begin{eqnarray}
\mathcal{A} = 
	\begin{pmatrix}
a_{11} & \cdots & a_{1n} \\
\vdots & \ddots & \vdots \\
a_{n1} & \cdots & a_{nn} \\
	\end{pmatrix}. \nonumber 
\end{eqnarray}
 The matrix $\mathcal{A}$ corresponds to an image data, and its component $a_{ij}$ represents the pixel value for  the $i$ th row and $j$ th column pixel. We set the size of image $n \times n$ here.  We also consider a matrix $\mathcal{C}$ as convolution filter given by
\begin{eqnarray}
\mathcal{C} = 
	\begin{pmatrix}
c_{11} & \cdots & c_{1m} \\
\vdots & \ddots & \vdots \\
c_{m1} & \cdots & c_{mm} \\ 
	\end{pmatrix}, \nonumber
\end{eqnarray}  
where $c_{ij}$ represents the pixel value for the $i$ th row and $j$ th column pixel. The size of the convolution filter is $m \times m$. Note that the convolution filter $\mathcal{C}$ is smaller than the image $\mathcal{A}$, i.e., $m < n$ .  We denote the convolution operation between $\mathcal{A}$ and $ \mathcal{C}$ by $\mathcal{A} \ast \mathcal{C}$. The output of  $\mathcal{A} \ast \mathcal{C}$ is a two-dimensional image.  First, we  stack the upper left corner of $\mathcal{A}$ and $\mathcal{C}$.  Then  the (1,1) component  of    $\mathcal{A} \ast \mathcal{C}$ ,  $(\mathcal{A} \ast \mathcal{C})_{1 1}$is calculated by $ \sum_{j=1}^{m} \sum_{k=1}^{m}  a_{jk} c_{jk}$.  If we reshape the stacked region of $\mathcal{A} $ and $\mathcal{C}$ to one-dimensional vectors $(a_{11}, \cdots , a_{1n}, \cdots , a_{n1}, \cdots a_{nn})$ and  $(c_{11}, \cdots , c_{1n}, \cdots , c_{n1}, \cdots c_{nn})$, respectively,   the convolution operation is able to be interpreted as the inner product of these vectors. This implies that the output of the convolution operation tells us whether the stacked area of $\mathcal{A}$ resembles that of $\mathcal{C} $.  Next, we move $ \mathcal{C}$ to the right by one pixel and calculate $(\mathcal{A} \ast \mathcal{C})_{1 2} = \sum_{j=1}^m \sum_{k=1}^m a_{j \, k+1} c_{jk} $. Similarly,  $(\mathcal{A} \ast \mathcal{C})_{p q }$ is computed by
	\begin{eqnarray}
(\mathcal{A} \ast \mathcal{C})_{p q} =\sum_{j =1 }^m \sum_{k=1}^m  a_{j+p-1 \, k+q-1} c_{j k}, \nonumber
	\end{eqnarray} 
where, $p$ and $q$ are integers between $1$ and $n-m+1$. When $p = n - m + 1$ is considered, the right end of $\mathcal{A}$ coincides with  that of $\mathcal{C}$. When $q = n - m + 1$ holds, the bottom edge of $\mathcal{A}$ coincides with  that of $\mathcal{C}$.  When the stride is 1 and no-padding case, the shape of  $ \mathcal{A} \ast \mathcal{C} $ is $ (n- m +1) \times (n -m +1)$. 
In Fig. \ref{fig:conv1} and Fig. \ref{fig:conv2}, examples of convolution operations are shown. In Fig. \ref{fig:conv1},  $\mathcal{A}$ is a square and we consider two $3 \times 3$ convolution filters. In Fig. \ref{fig:conv2},  $\mathcal{A}$ is a cross and convolution filters are same as in Fig. \ref{fig:conv1}. Looking at the result of the convolution operation, we are able to know where patterns of $\mathcal{C}$  are contained in $\mathcal{A}$ and this helps us predict the shape of  objects in $\mathcal{A}$. The channel number of output of the convolution layer is the number of convolution filters.  In Fig. \ref{fig:conv1} and Fig. \ref{fig:conv2}, the channel number of output images is two.

\begin{figure}[htb]
 \centering
   \includegraphics[width=100mm]{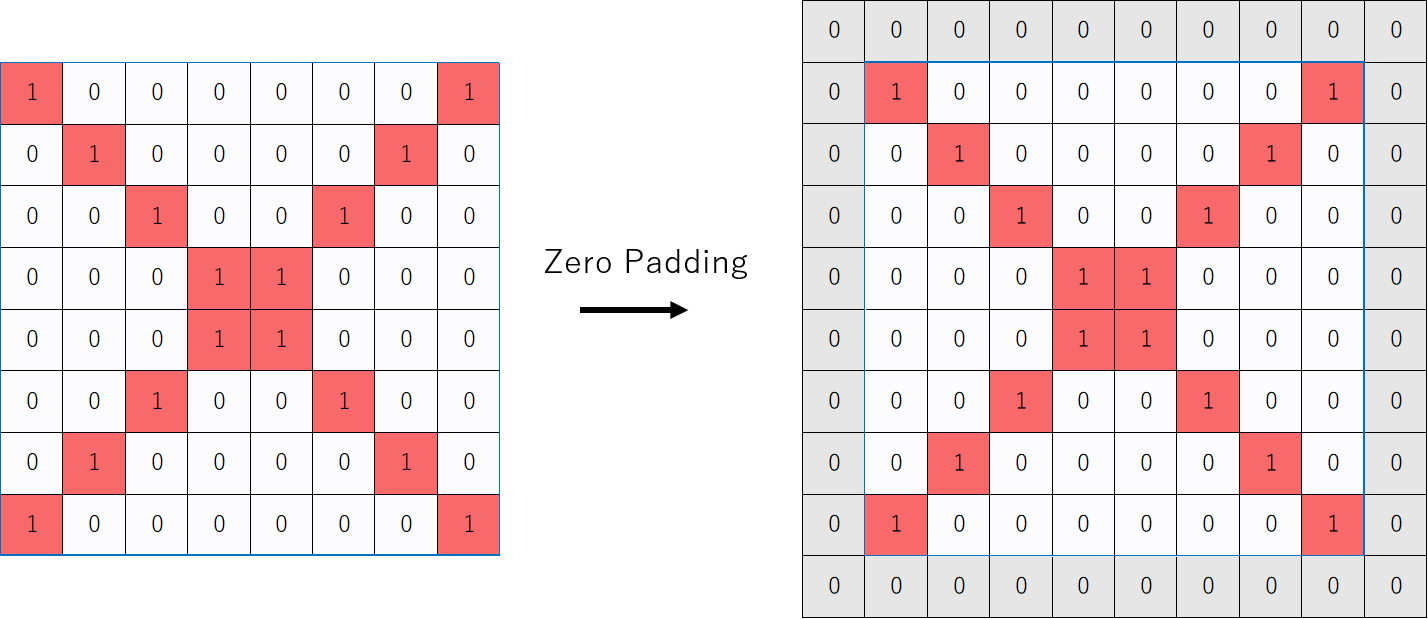}
 \caption{An example of zero padding. The size of original image (left figure) is $8 \times 8$. If we apply a convolution operator using a filter whose size is $3 \times 3$ and we want  the size of the output image of the convolution operator to be $8 \times 8$, we have to enlarge the input image to the $ 10 \times 10$ size. Values of margin is 0 in the zero padding. }
 \label{fig:zero_pad}
\end{figure}

If we would like  the output image of the convolution layer to be $n \times n$, we apply padding to $\mathcal{A}$. In the padding, we enlarge $\mathcal{A}$ so that the size of $\mathcal{A}$ is $ (n+ m - 1) \times (n+m-1) $ and pad values in blank spaces between the enlarged image and the original image. For example,  in the zero padding  case, we pad 0 in the margin like Fig. \ref{fig:zero_pad}.  
\noindent
In the above explanation, we consider the stride is 1 for simplicity. That is to say, we move the convolution filter one by one. In general, the stride needs not be set to 1. If we take the stride to 2,  $\mathcal{C}$ moves two squares.

\begin{figure}[H]
 \centering
   \includegraphics[width=130mm]{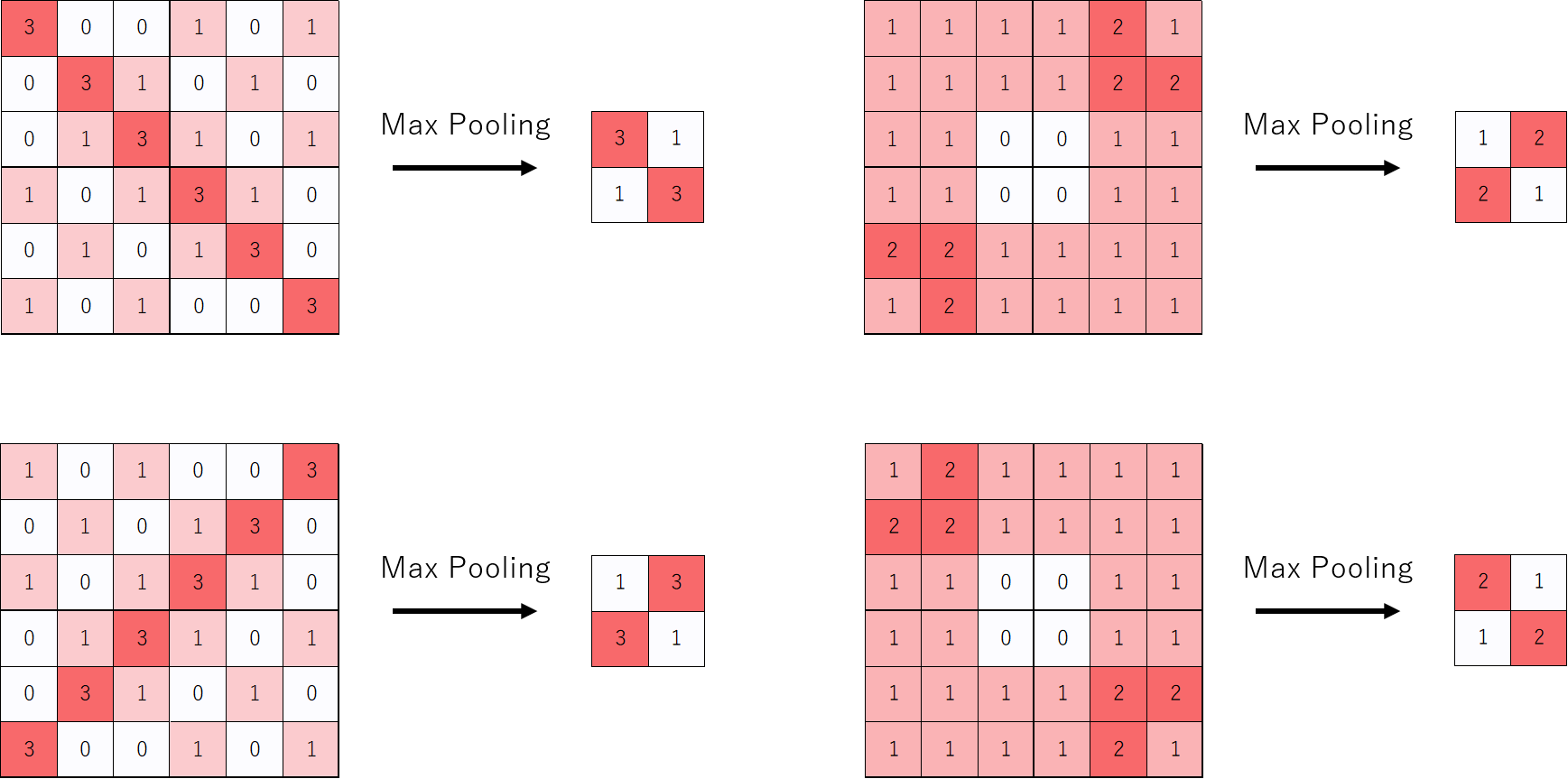}
 \caption{The example of max Pooling.}
 \label{fig:pool}
\end{figure}

\begin{figure}[H]
 \centering
   \includegraphics[width=130mm]{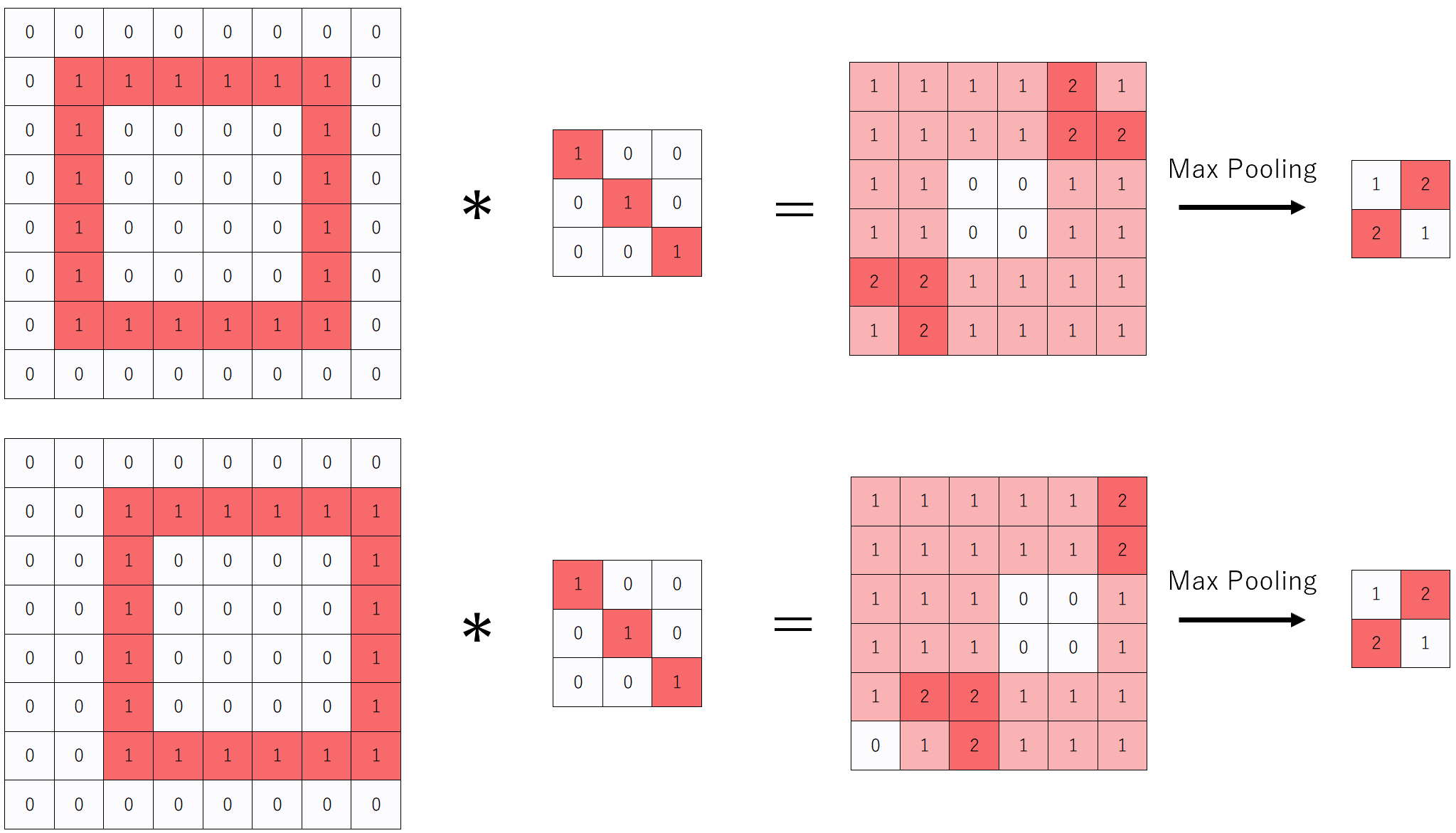}
 \caption{Robustness of the parallel shift: positions of red square are different in two original images. The results of the convolution operation are also different. However, outputs are same after applying the pooling and it gives a DL model parallel invariance. }
 \label{fig:shift}
\end{figure}

\begin{figure}[H]
 \centering
   \includegraphics[width=110mm]{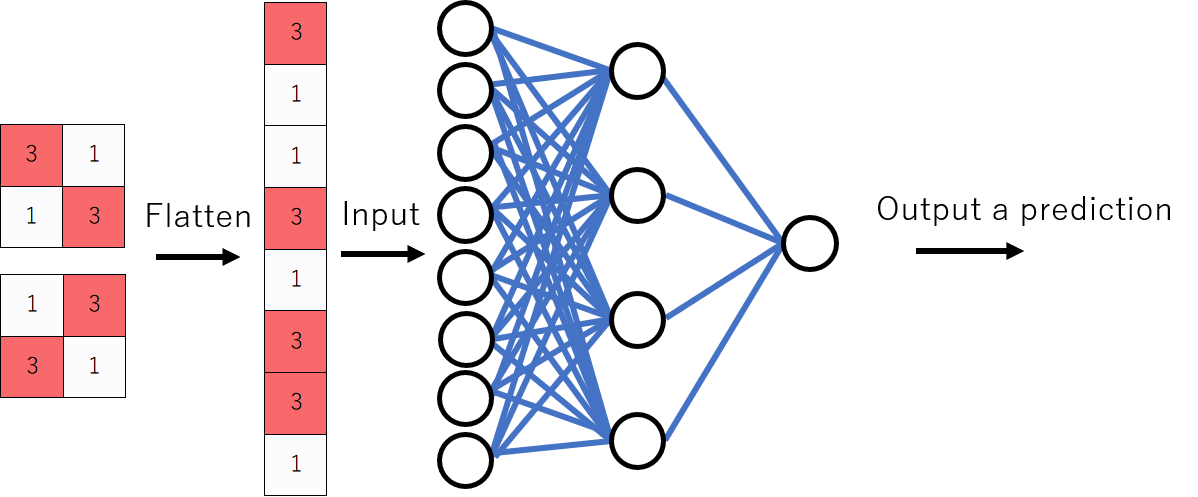}
 \caption{Neural Network: Finally, we flatten the output of convolution and pooling layers and input it to a neural network. Neural network returns a prediction of a target label.  }
 \label{fig:nn}
\end{figure}

Next, we explain the function of the pooling layer.  The pooling layer downsizes images and reduces the amount of data. For example,  we adapt the max-pooling to the outputs of the convolution operation, which are illustrated in Fig. \ref{fig:conv1} and Fig. \ref{fig:conv2}, results of the pooling operation are depicted in Fig. \ref{fig:pool}. In Fig. \ref{fig:pool}, we divide images into $ 3 \times 3$ pixels and  calculate  maximum values of each area.  In this case,  the numbers of rows and columns of divided areas are both two, and the shapes of output images of pooling operations are $2 \times 2$. 
The pooling layer has an important role. It gives the robustness of the parallel shift of objects in pictures.  An example is shown in Fig. \ref{fig:shift}. The upper-left image and the lower-left image both contain the square, but the locations of the squares are different.  Therefore results of the convolution operation, where the convolution filter are same in both cases,  are also different. However, when we want to know whether the image includes a square or not,  the prediction of DL for both images should be same.  Thus the DL model must have the robustness of the parallel shift of objects.  The pooling layer gives the DL model the robustness of the parallel shift, as illustrated in Fig. \ref{fig:shift}.
Finally, we explain the role of standard neural networks (NN).  The input of neural networks is a vector that is generated by flattening images after convolution layers, and pooling layers act as shown in Fig. \ref{fig:nn}.  The output of NN is a prediction.  In neural networks, linear transformations described by weight matrices and non-linear transformations represented by activation functions like the sigmoid function are alternately operated on an input vector.  NN is able to approximate a complicated non-linear function by adding hidden layers.

\subsection{Flow of DL}
In this subsection, we explain the flow of DL. First we prepare a data set. Then we split the data  into training data, validation data and test data. Training data are used for training a DL model. In CNN, values of convolution filters and weights of NN $w$  are parameters to be learned. Validation data are utilized to check whether the DL model overfits for the training part. The overfitting is the phenomenon that the DL model's generalization ability loses. The test data are used to evaluate the model performance. 

Next, we introduce a loss function for the DL. We denote by $x^{(i)}$ and $y^{(i)}$ the $i$ th feature vector and the target label, respectively.  Let us represent the DL prediction for the $i$-th data as $\bar{y} (x^{(i)},w)$. It is the function that quantifies how close the prediction and the expected answer are. For the regression problems, the mean square error (MSE) 
\begin{equation}\label{loss}
    E=\frac{1}{n}\sum^n_{i=1}\big(\bar{y}(x^{(i)},w)-y^{(i)}\big)^2,
\end{equation} 
is frequently used as the loss function, where $n$ is the number of data. On the other hand, in the classification problems, one of the most common loss functions can be the cross entropy,
\begin{equation}
    E=-\frac{1}{n}\sum_{i=1}^n \sum^M_{c=1} y_c(x^{(i)},w) \log P_c(x^{(i)},w),
\end{equation}
where $M$ is the total number of classes, $y_c(x^{(i)},w)$ is the target one-hot vector, which takes 1 for a target class or 0 for the rest, and $P_c(x^{(i)},w)$ is the probability when the DL model predicts that  the $i$-th data 's target label is $c$.  The final prediction of DL $\bar{y}(x^{(i)},w)$ will be the class that has the highest probability $\bar{y}(x^{(i)},w) = {\rm argmax}_{c} P_c(x^{(i)},w)$. 

The procedure of DL is the following. First we give training data, validation data and  random initial values of $w$ to a DL model. Second, DL calculates predictions $\bar{y}(x^{(i)},w) $ and a loss function $E$ for training data and validation data . Third, we calculate the gradient of the loss function for training data and update parameters $w$ to decrease the loss function. Specifically, we used the Adam method for the parameter optimization\cite{Adam} in simulations described in later sections.  

In Adam, the learning parameters $w$ are updated by the following equations:
\begin{eqnarray}
v_t &=& \beta_1 v_{t-1} + (1-\beta_1) \partial_w E , \\
s_t &=& \beta_2 s_{t-1} + (1-\beta_2) (\partial_w E)^2, \\
w_t &=& w_{t-1} - \alpha \frac{v_t}{\sqrt{s_t + \epsilon}},
\end{eqnarray}
where, $t$ is the index representing epochs and $w_t$ are parameters at the $t$ epoch. $\beta_1$ ,$\beta_2$, $\alpha$ and $\epsilon$ are hyperparameters whose values are explicitly defined by the users of DL. $\alpha$ must be positive and called learning rate. $\beta_1$ and $\beta_2$ must take values between $0 < \beta_1,\beta_2 <1$. $v_t$ and $s_t$ are ancillary variables to compute the new parameter $w_t$. The initial value of these parameters is $v_0 = s_0 = 0$.  

Fourth, we repeat calculating the loss function and updating parameters. The number of times of updating parameters is called epoch. The loss function for train data  generally decreases as the epoch increases. The loss function for validation data also decreases but from some epoch, it turns to increase. This means that the overfitting happens. So, we stop updating parameters when the loss function for validation data is  remarkably  increasing. Finally, we give test data to the DL model and evaluate the performance like the accuracy and the loss function for test data. The flow of DL is illustrated in AL. \ref{fig:flowchart}.

\begin{figure}[!t]
    \begin{algorithm}[H]
    \caption{Flow of DL}
    \label{fig:flowchart}
    \begin{algorithmic}[1]
    \STATE Prepare dataset $\mathcal{D}$ and split it to training data $\mathcal{D}_{train}$ , validation data $\mathcal{D}_{vali}$ and test data $\mathcal{D}_{test}$.
    \STATE Decide which model we use and values of hyperparameters and initialize learning parameters $w$.
    \WHILE{The loss function $E$ for $\mathcal{D}_{vali}$ continuously decreases,}
    \STATE Calculate the loss function $E$ for $\mathcal{D}_{train}$ and $\mathcal{D}_{vali}.$
    \STATE Update parameters $w_t \leftarrow w_{t-1}$ using  optimization methods like Adam.
    \ENDWHILE
    \STATE Evaluate the model performance like the accuracy and the loss function using $\mathcal{D}_{test}$.
    \end{algorithmic}
    \end{algorithm}
    
\end{figure}
\section*{Appendix B}
In this appendix, we prove the solution of the equation of motion of free particle in 2+1 AdS spacetime.
The metic is given by 
\begin{eqnarray}
ds^2 = -(1+\lambda^2 r^2) dt^2 + \frac{dr^2}{1+\lambda^2 r^2} + r^2 d\phi^2.
\end{eqnarray}
The first derivative of the proper time $\tau$ must satisfy the following relation:
\begin{eqnarray}
(1+ \lambda^2 r^2) \left( \frac{dt}{d\tau}\right)^2 - \frac{1}{1+\lambda^2 r^2} \left( \frac{dr}{d\tau}\right)^2 - r^2\left( \frac{d\phi}{d\tau}\right)^2 = 1. \label{eq:proper_time}
\end{eqnarray}
From the energy and the angular momentum conservation laws, we find 
\begin{equation}
\frac{1}{2} \left( \frac{dr}{d\tau}\right)^2 + \frac{1}{2} \left(\lambda^2 r^2 + \frac{L^2}{r^2} \right) = E,\quad
r^2 \frac{d\phi}{d\tau}  = L. 
\label{eq:energy_conserve}
\end{equation}
From Eq. (\ref{eq:energy_conserve}), we obtain $r(\tau)$ by integrating 
\begin{eqnarray} 
\tau = \pm \int \frac{dr}{2E - \lambda^2 r^2 - L^2/r^2}=\pm \frac{1}{2} \int \frac{du}{\sqrt{2Eu - \lambda^2 u^2 - L^2}},
\end{eqnarray} 
where introduce $u$ defined by $u = r^2$.
Because the argument of square root must be non-negative, 
$E \geq \lambda L$ 
has to be satisfied. 
Next, we define $v$ as $ v = u - E/\lambda^2$. The integral becomes 
\begin{eqnarray}
\tau  = \pm \frac{1}{2\lambda}\arcsin\left[\frac{\lambda}{\sqrt{E^2/\lambda^2-L^2}} v \right].
\end{eqnarray}
Then we obtain the relation between $r$ and $\tau$ as follows:
\begin{align}
&\tau = \pm \frac{1}{2\lambda} \arcsin\left[\frac{\lambda}{\sqrt{E^2/\lambda^2-L^2}} \left(r^2 - \frac{E}{\lambda^2} \right) \right] + \tau_0, \\
\pm&  \sin \left[2 \lambda (\tau -\tau_0)\right] = \frac{\lambda}{\sqrt{E^2/\lambda^2-L^2}} \left(r^2 - \frac{E}{\lambda^2} \right),\\
&r^2 = \frac{1}{\lambda^2} \{ E \pm \sqrt{E^2-\lambda^2 L^2}\, \sin [2\lambda (\tau -\tau_0)]\}. \label{eq:r_squared}
\end{align}
We are now considering periodic orbits images, $\tau_0$ can be set to 0 and we can choose the negative sign. Finally, because $r$ must be non-negative, the solution $r(\tau)$ is given by
\begin{eqnarray}
r = \frac{1}{\lambda} \sqrt{ E - \sqrt{E^2-\lambda^2 L^2} \sin [2\lambda \tau]}. \label{}
\end{eqnarray}

Next, we calculate the solution of $\phi$. For later convenience,  we recall the following formula of integration:
\begin{eqnarray}
\int  \frac{dx}{a -b \sin2x} = \frac{a}{\sqrt{a^2 - b^2}} \arctan \left[ \frac{a}{\sqrt{a^2 - b^2}} \left( \tan x - \frac{b}{a} \right)\right] . \label{eq:int_formula}
\end{eqnarray}
Using  Eq. (\ref{eq:int_formula}), we derive the solution of $\phi$. From Eq. (\ref{eq:energy_conserve}), we obtain
\begin{eqnarray}
\phi = \int d\tau  \frac{L}{r^2} 
= \int d\tau  \frac{L\lambda^2}{ E - \sqrt{E^2-\lambda^2 L^2} \sin [2\lambda \tau]}.
\end{eqnarray}
 Substituting $ x = \lambda \tau $, $a = E $ and $b = \sqrt{E^2-\lambda^2 L^2}$, we find
\begin{eqnarray}
\phi =  \arctan \left[\frac{E}{\lambda L} \left(\tan \lambda \tau - \sqrt{1 - \left(\frac{\lambda L}{E}\right)^2} \right)  \right],
\end{eqnarray}
here, we used $\tau \geq 0$. Looking at Eq.(\ref{eq:r_squared}), there is a freedom of sign of $\phi$. Thus, the solution of $\phi$ is given by 
\begin{eqnarray}
\phi = \pm  \arctan \left[\frac{E}{\lambda L} \left(\tan \lambda \tau - \sqrt{1 - \left(\frac{\lambda L}{E}\right)^2} \right)  \right] + \phi_0 \label{eq:phi_sol}.
\end{eqnarray}
Finally, we compute the solution of $t$ using Eq. (\ref{eq:proper_time}).
From Eq. (\ref{eq:proper_time}) and  Eq. (\ref{eq:energy_conserve}), we obtain
\begin{eqnarray}
\left(\frac{dt}{d\tau}\right)^2 &=& \frac{1}{1+\lambda^2 r^2} + \frac{1}{(1+\lambda^2 r^2)^2} \left(\frac{dr}{d\tau} \right)^2 + \frac{r^2}{1+\lambda^2 r^2} \left(\frac{d\phi}{d\tau} \right)^2  \notag \\
&=& \frac{1}{(1+\lambda^2 r^2)^2}  \left[ 1 + 2E +\lambda^2 L^2 \right] \notag \\
&=& \frac{ 1 + 2E +\lambda^2 L^2}{(1 + E - \sqrt{E^2-\lambda^2 L^2} \sin [2\lambda \tau])^2}.
\end{eqnarray}
Then, substituting  $a = 1+E$ and $b = \sqrt{E^2-\lambda^2 L^2} $ in the formula \eqref{eq:int_formula}, we find that
\begin{eqnarray}
t &=& \int d\tau \frac{\sqrt{1 + 2E +\lambda^2 L^2}}{1 + E - \sqrt{E^2-\lambda^2 L^2} \sin [2\lambda \tau]}\notag  \\
&=&  \frac{1}{\lambda} \arctan \left[  \frac{1}{\sqrt{1+2E + \lambda^2 L^2}} \left((1+E)\tan \lambda \tau -  \sqrt{E^2-\lambda^2 L^2}   \right)  \right] .
\end{eqnarray}
Therefore, the solution of $t$ is given by 
\begin{eqnarray}
t = \frac{1}{\lambda} \arctan \left[  \frac{(1+E)\tan \lambda \tau -  \sqrt{E^2-\lambda^2 L^2} }{\sqrt{1+2E + \lambda^2 L^2}}\right]+ t_0 .
\end{eqnarray}


\begin{thebibliography}{9}
\bibitem{comment} At a later stage, we prepare the image with 70 particle trajectories, the accuracy increases to about 96\% .
\end{thebibliography}
\end{document}